\documentclass[letterpaper]{article}
\usepackage[utf8]{inputenc}
\usepackage{amsmath}
\usepackage{amssymb}
\usepackage{amsthm}
\usepackage{txfonts}
\usepackage{authblk}
\usepackage{breakcites}
\usepackage{hyperref}
\usepackage{hyphenat}
\usepackage{color}
\usepackage{tikz}
\usepackage[scr]{rsfso}
\usepackage{tabularx}
\newtheorem{theorem}{Theorem}[section]
\newtheorem{lemma}[theorem]{Lemma}
\newtheorem{example}[theorem]{Example}
\newtheorem{definition}[theorem]{Definition}
\newtheorem{corollary}[theorem]{Corollary}
\newtheorem{proposition}[theorem]{Proposition}
\newtheorem{remark}[theorem]{Remark}

\usepackage{graphicx}
\usepackage[all]{xy}

\renewcommand{\phi}{\varphi}

\newcommand{\val}[1]{[\![{#1}]\!]}

\newcommand{\marginnote}[1]{\marginpar{\raggedright\tiny{#1}}}
\newcommand{\wccolor}{\color{red}}
\newcommand{\wcprefix}{W: }
\newcommand{\wcnote}[1]{{\wccolor \wcprefix #1 }}
\newcommand{\wcmnote}[1]{\marginnote{\wcnote{#1}}}

\usetikzlibrary{arrows}

\title{The Tiglath Pileser principle and models of persuasion}
\author{Willem Conradie,  Krishna B.~Manoorkar,  Alessandra Palmigiano, Apostolos Tzimoulis,  Nachoem M.~Wijnberg}
\date{\today}

\begin{document}

\maketitle
\begin{abstract}

This paper develops a formal framework for modelling how agents extract reliable information from narratives and communicative interactions in which speakers may strategically distort the truth. Building on  the codex-based models of persuasion introduced by Rubinstein and Glazer, we shift the focus from codices to their associated guidance relations, and generalize these to a broader class of \emph{Tiglath-Pileser relations}, which need not be  induced by any underlying codex. We use Tiglath-Pileser relations to formally represent the application of what Baruch Halpern terms the  {\em Tiglath-Pileser principle} for extracting reliable information, based on the idea that, while they might be systematically biased, reported claims must maintain a connection with the truth. 
%
%
We provide a set-theoretic and axiomatic analysis of these relations, identifying structural conditions under which they support successful applications of the Tiglath-Pileser principle in persuasion settings. Within this framework, Rubinstein and Glazer's persuasion scenarios and instances of magical implementation are reformulated in relational terms, thereby extending their scope beyond one-shot interactions with common knowledge of agents’ goals.
A central feature of the present approach is its dynamic perspective: it accounts for how Tiglath-Pileser relations  are formed, revised, and strategically manipulated over the course of a given narration or communicative interaction, giving rise to higher-order reasoning about the reliability and intentions of information sources.
This framework is applied to two case studies; in particular, we show how the proposed machinery can formally reconstruct inferential patterns underlying historiographical analysis, such as Baruch Halpern’s interpretation of the biblical account of King David.
The methodology adopted in this paper integrates tools from logic, epistemic reasoning, and game theory, contributing a formal perspective on the dynamics of belief formation under strategic information transmission, and providing a foundation for analysing how agents can approximate the truth in environments shaped by partial information, bias, and manipulation. 

    
    

    
\end{abstract}
\tableofcontents

\section{Introduction}
A core question in historiography (as in many other academic and non academic fields) is: how much of a text (or of a story as is told in a text or other medium) can we believe? 
In his analysis of the story of King David as it is told in the Book of Samuel and 2 Kings \cite{halpern2003david}, the biblical historiographer  Baruch Halpern  
suggests (cf.~\cite[Part III, Ch.~5]{halpern2003david}) that this text\footnote{
From \cite[Ch 3 p.~57]{halpern2003david}: the basic question is, why do we have this text? What was its purpose? Who wrote it? The answer to the riddle is straightforward: most of David’s story was written during Solomon’s reign, and the object was both to glorify the founder of the dynasty and to advance Solomon’s political position.} can be analysed using the same basic interpretive strategy which applies to the Assyrian annals of Tiglath-Pileser,  
in which  
 ``the torching of a grain field is the conquest of a whole territory beyond it. A looting raid becomes a claim of perpetual sovereignty. But this does not mean that campaigns can be confected. The 
technique is that of putting extreme spin on real events.'' Halpern continues (ibid.):
\begin{quote}
    
Interpreting such literature demands only a simple rule, the Tiglath-Pileser 
principle. The question is, what is the minimum the king might have done to 
lay claim to the achievements he publishes? Looting a town? He shoplifted a 
toothbrush from the local drug store. Ravaging the countryside? Perhaps he 
trampled crops near a farmstead. Receiving submission from distant kings in 
lands one hasn’t invaded? A delegation arrived to inaugurate diplomatic relations. Each small mark of prestige becomes the evidence for a grand triumph. 
[...] 
The point is, 
such figures cannot make claims without any basis in fact [...] 
But it is always a temptation to paint one’s achievements in the best of all possible lights. To correct 
for this tendency, one has merely to imagine what insignificant action would 
produce the claim that ``I conquered Egypt” (a raid on a border station in the 
Sinai). The same political hermeneutic applies reasonably well today. 
\end{quote}

Thus, the Tiglath-Pileser principle is grounded in the assumption that the given narration is anchored in truth (``this does not mean that campaigns
can be confected'', ``such figures
cannot make claims without any basis in fact''), because the  narrator must take into account what knowledge  his original audience has that cannot be contradicted. Moreover, the Tiglath-Pileser principle  provides a way of attaining a more accurate approximation of the truth, by taking into account both the original audience's knowledge of the actual circumstances (which constrain the narrator to not make claims that would be immediately disbelieved), and  the narrator's agenda, and thus correcting for the tendency of this agenda to `spin the truth'. Building upon this ‘simple rule’ enables Halpern to reach the following surprising conclusions in respect to 
King David's story  \cite[Afterword, p.~479]{halpern2003david}:

 \begin{quote}
    [...] 
    the only murder in the books of Samuel of which [David] was probably innocent is the one murder of which he stands accused in the apology. 
%
%
[...] 
At the end of his life, David was certainly betrayed by one of his harem women, Bathsheba, and his foreign mercenary bodyguard. Instead of his heir, Adonijah, they installed Solomon, hitherto known as the son not of David but of Uriah the Hittite, on the throne. 
[...] this second king of Judah quickly cleaned house, erasing his rivals, and probably all of David's sons, from the scene. 

 \end{quote}
This paper presents a framework which, {\em inter alia}, can formally represent and trace the steps from the Tiglath-Pileser principle to Halpern's conclusions. This framework builds on the theory developed by Rubinstein and Glazer in a series of papers \cite{glazer2012model, glazer2014complex, glazer2024magical} in which several types of situations are analysed in which one or more agents (the Speakers) wish to {\em persuade} another agent, the Listener (who plays  the principal's role), \footnote{The term ``agent'' is used here in two partially overlapping but historically distinct senses. In the Rubinstein–Glazer framework, inherited from the principal–agent literature in economics \cite{ross1973economic}, an agent is understood relationally as a party acting on behalf of, or strategically interacting with, a principal under conditions of informational asymmetry and potentially divergent interests. By contrast, in the epistemic-logic and AI tradition \cite{fagin2004reasoning} which also informs the present framework, agents are typically construed as autonomous epistemic subjects characterised by informational states, inferential capacities, strategies, and action possibilities. The present approach combines these perspectives by modelling agents both as bearers of epistemic attitudes and as strategically situated participants in persuasion processes. In this respect, both the Speaker(s) and the Listener are regarded as ``agents''.} to make a decision according to their desiderata, e.g.~to accept a particular request, to award a prize to them, or to believe a certain claim they make to be true. The Listener's decision depends on what they conclude about a certain state of affairs, which can either concern the (particular) Speaker (e.g.~whether the Speaker satisfies certain requirements) or some other facts about the world.    Each Speaker knows the true  state of affairs, whereas the Listener knows only the set $S$ of the possible states of affairs (and possibly some information about the Speaker's goals and strategies). Each Speaker wishes to persuade the Listener  regardless of the true state of affairs. Hence, the Speakers can lie, and will do so whenever declaring the true state of affairs would not favour them. However, the key feature of the scenarios in \cite{glazer2012model,  glazer2024magical} is that the Speakers' lies are {\em constrained} by a particularly strong bounded rationality assumption that  anchors their lies to the truth. Hence, in certain cases, the Listener can engineer the enquiry in such a way that they can approximate the true state of affairs to varying degrees of accuracy. To formally represent this idea, Rubinstein and Glazer introduce the notion of {\em codex}, a finite set $L$ of constraints, expressed as formulas of classical propositional logic, which the Speaker's  declared state is expected to satisfy. The specific way in which the Speakers' lying ability is constrained is formalized  by the {\em guidance relation} ${\to_L}\subseteq S\times S$ associated with $L$, where 
$s\to_L s'$ if, whenever $s$ is the true state, the Speaker could declare $s'$ in an attempt to satisfy codex $L$.  

Guidance relations are perspicuous formal representations of the Tiglath-Pileser principle in action: by knowing the Speaker's guidance relation,  the  Listener is able to conclude that the true state must  belong to the preimage of  $s'$ under $\to_L$,  whenever the Speaker declares $s'$ as the `true state'.   Depending on the properties of the guidance relation, the Listener is able to gauge the information needed for their decision.

We  introduce a  class of relations, referred to as {\em Tiglath-Pileser relations}, which  generalize  guidance relations, and need not arise from codices. 
We pursue an axiomatic development of the properties of Tiglath-Pileser relations required for the successful application of the Tiglath-Pileser principle. 
We introduce a logic-based framework for the analysis of scenarios in which the Listener's approximations of the Speaker's Tiglath-Pileser relation are dynamically formed over the course of extended interaction, and  
 apply it to two case studies.

The framework introduced in the present paper differs from that of \cite{glazer2012model, glazer2014complex, glazer2024magical} in important respects. Firstly, while in \cite{glazer2012model,  glazer2024magical}, the constraints on the Speakers' ability to lie are purely a result of their bounded rationality, we explore situations in which the background knowledge of the Listener imposes further constraints. Secondly, we are interested in exploring  more open-ended scenarios than those  considered in  \cite{glazer2012model, glazer2014complex, glazer2024magical}, where the latter mostly concern the structuring of 
selection procedures designed to elicit the truth in situations which can be modelled as one-off interactions, and the goals of the Speaker(s) are common knowledge among all agents. In contrast, we are interested in accounting for possibly extended interactions among agents, in which the goals of the Speaker(s) might not be explicitly known, and in fact, the Speaker(s) might have an interest in concealing them both from their primary intended audience, and from any agent who, playing the `historian's role', tries to interpret the whole communication.
Thirdly, in contrast to \cite{glazer2012model, glazer2014complex, glazer2024magical}, we take a dynamic perspective on how guidance relations arise in the Listener's mind, and how this process can be manipulated by strategic Speakers to feed the Listener a false relation.\footnote{A classic example of such manipulation is that of a conjurer or stage magician, who tricks the audience into believing that he is cheating in a certain way, thereby blind sighting them to the actual trick. Certain strategies of con artists are another example.}
Fourthly,  in \cite{glazer2012model,  glazer2024magical}, the Speakers are assumed to ``not think {\em strategically}, i.e.~not [to] take into consideration the other
party’s actions'', but rather act as ``problem solvers'', e.g.~by (only) making sure that their declarations verify the codices they have been assigned; in contrast, we structure the analysis of the present paper according to  a hierarchy  of levels of strategic/cognitive  \cite{stahl1994experimental,nagel1995unraveling,camerer2004cognitive} of the agents involved, the most basic level of which is that of the ``non-strategically thinking'' agents considered in \cite{glazer2012model,  glazer2024magical}.
While, like \cite{glazer2012model,  glazer2024magical}, we avoid an explicit game-theoretic treatment, this approach sets the stage for a game-theoretic development of signalling games with uncertain information and players acting at different levels of the strategic/cognitive hierarchy. 
In this perspective, this paper contributes  novel  insights about the formal modelling of \emph{trust}  based on  inferring the motivations and persuasion strategy of  other  agents, rather than purely on their track-record of truth-telling in previous rounds of a game \cite{KREPS1982253,atakan2012reputation}.   

The framework developed here has direct applications to detecting and managing disinformation. In adversarial information environments, strategic actors rarely invent entirely false narratives;  instead, they often exploit, selectively amplify, and distort fragments of existing information to create misleading impressions \cite{wardle2017information}. By modelling propagandists as  strategic Speakers constrained by the information their audience possess, our framework enables analysts, algorithms, and fact-checkers (Listeners) to attain a reliable approximation of the truth embedded within manipulated narratives. The present treatment of extended interactions and higher-order reasoning also models how sophisticated campaigns shape perceptions of both events and source motives.

Another area of application concerns AI safety and the governance of agentic systems. As AI agents become more capable, risks such as sycophancy and deceptive alignment become increasingly important \cite{amodei2016concrete}. This framework makes it possible to model AI systems as strategic Speakers interacting with human Listeners, and hence it can be used by supervisors to audit, probe, and constrain potentially deceptive behaviour. Such an approach aligns with emerging work on scalable oversight, interpretability, and evaluations designed to detect undesirable behaviours in advanced AI systems \cite{bai2022constitutional,christiano2017deep}.

\paragraph{Structure of the paper.} In Section \ref{sec: preliminaries}, we give a detailed summary and analysis of the formalisms introduced in \cite{glazer2012model, glazer2024magical}; in Section \ref{Sec:Tiglath-Pileser:Relations}, we further develop and refine this formal framework, with a particular focus on the relationship between codices and guidance relations; in Section \ref{sec:TP relations and game-theoretic}, we recast this formalism in a purely set-theoretic way; in this formal environment, the analysis is structured along a hierarchy of levels of strategic/cognitive awareness of the agents involved, successive levels of which correspond to deeper levels of sophistication in the agents' reasoning about other agents' motivations and Tiglath-Pileser relations; in Section \ref{Sec:Dynamically:Formed:Relations}, we introduce a logic-based framework which  accounts for how codices and their associated Tiglath-Pileser relations are dynamically generated in the mind of the Listener in the context of a narration which unfolds over several episodes, how these formal structures govern the process of  formation of reliable beliefs in the Listener's mind, how this process can be manipulated by strategic Speakers, and how the Listener can outmanoeuvre the Speaker in these manipulative attempts. This framework is applied to case studies in   Sections \ref{sec: granpa wijnberg} and \ref{sec: king david}, the latter  section collecting the formal account of Baruch Halpern's analysis of King David's story. We conclude in Section \ref{sec: conclusions}.

\section{Preliminaries}
\label{sec: preliminaries}

In this section, we review the formalisms introduced in \cite{glazer2012model, glazer2024magical}, where they are embedded in extensive motivating case studies of persuasion scenarios. Our presentation abstracts from the latter, while highlighting and developing the  mathematical and formal aspects, preparing the ground for the introduction of the generalization to Tiglath-Pileser relations in Section \ref{Sec:Tiglath-Pileser:Relations}.    


\subsection{Basic formalism of `A model of persuasion'}\label{sec: basic formalism}

For any (finite) set $\mathsf{Prop}  = \{v_1, \ldots, v_K\}$ of propositional variables\footnote{Throughout this paper, we assume that the set $\mathsf{Prop}$ is linearly ordered, with the order  given by their order of appearance from left to right. This order determines the order of the coordinates in the tuples of $2^\mathsf{Prop}$.} with $K \geq 2$,
a \textit{profile}, or {\em state}\footnote{In what follows, we typically reserve the term `profile' to maps $s: \mathsf{Prop}\to \{0,1\}$, and let $S = 2^\mathsf{Prop}$ denote the domain of profiles, while we refer to  the elements of $S$ as `states' when $S$ is an arbitrary set.}, is a map $s: \mathsf{Prop}\to 2 = \{0, 1\}$; we let $S$ denote the set of states and write $s\models v$ iff  $s(v) = 1$. 




In \cite{glazer2012model},  leader-follower scenarios are analysed in which a Listener announces and commits to a \textit{codex} $L$, which is a logically  consistent set of  propositional formulas in the language of $V$ that the profile presented by a Speaker must satisfy in order for the Speaker’s request to be accepted. Then, the Speaker (who knows the true profile) announces a profile that may or may not be the true one. 


Each formula (rule) $\varphi$ in the codex is required to have the following syntactic shape:
\[
\bigwedge_{y \in Y} \varphi_y \to \varphi_x,
\]
where $Y$ is a nonempty subset of $\mathsf{Prop}$, $x \in \mathsf{Prop} \setminus Y$, and each $\varphi_v$ is either $v$ or $\neg v$ (the negation of $v$) for any $v\in \mathsf{Prop}$. For example, the proposition $v_4 \land \neg v_1 \to v_3$ can be a rule in a codex, but $v_1 \to \neg v_1$ cannot. For every formula $\varphi$ and every set of formulas $\Gamma$, let $Var(\varphi)$ denote the set of variables occurring in $\varphi$ and let $Var(\Gamma) = \bigcup_{\varphi\in \Gamma} Var(\varphi)$. A rule $\varphi$ is {\em complete} if $Var(\phi) = \mathsf{Prop}$.

For any  rule $\varphi = \bigwedge_{y \in Y} \varphi_y \to \varphi_x$, we denote its antecedent and consequent as follows:
\[
a(\varphi) = \bigwedge_{y \in Y} \varphi_y \quad \quad \text{and }\quad \quad z(\varphi) = \varphi_x.
\]
%
Codices are required to be not only  satisfiable but also {\em coherent}, i.e.~for all $\varphi, \psi\in L$, if $a(\varphi)$ and $ a(\psi)$ are jointly satisfiable, then so are $z(\varphi) $ and $ z(\psi)$.

\begin{definition}[Guidance relation]
\label{def:guidance}
    Let $S = 2^K$. The {\em guidance relation} $\to_L$ associated with any codex $L$ is a binary relation ${\to_L}\subseteq S\times S$ defined as follows: $s\to_L s'$  
    iff for every variable $v\in \mathsf{Prop}$ such that $s'(v) \neq s(v)$,  a rule $\varphi \in L$ exists such that:

\begin{enumerate}
\item $z(\varphi)\in \{v, \neg v\}$;
    \item $s \models a(\varphi)$ and $s' \models a(\varphi)$, and
    \item $s' \models z(\varphi)$. 
\end{enumerate}

\end{definition}
Hence, if $s\to_L s'$, then there exists at least one rule $\varphi\in L$ s.t.~$s\not\models \varphi$ and $s'\models \varphi$, where the latter satisfaction is {\em non-trivial}, since  $s'\models a(\varphi)$. 
Codices $L$ consisting only of complete rules force their associated guidance relations to be  subsets of the {\em neighbourhood} relation $N\subseteq S\times S$ s.t.~$s N s'$ iff  some $x\in V$ exists s.t.~$s(v) = s'(v)$ for any $v\in V\setminus\{x\}$. For any two such neighbors $s$ and $s'$, consider the following complete rule \begin{equation}
\label{eq:phi-s-s-prime}
    \varphi(s, s') := \bigwedge_{v \in \mathsf{Prop} \setminus \{x\}} \varphi_v \to \varphi_x.
\end{equation} 
Thus by construction, $s$ is the only profile which falsifies $\varphi(s, s')$, since every other profile would either verify its conclusion or falsify its premises, and $s'$ is the unique profile which {\em nontrivially} satisfies  $\varphi(s, s')$, since $s'\models a(\varphi(s, s'))$. Moreover, $s \to_L s'$ for any codex $L$ that contains $\varphi(s, s')$.
For any codex $L$, let 
\[
T(L) : = \{ s\in S \mid s \models \varphi \text{ for all } \varphi \in L \} \quad \text{ and }\quad P(L) = \{s\in S \mid s \rightarrow_L s' \text{ for some } s' \in T(L)\}.
\]
That is, $T(L)$ is the set of profiles that, whenever declared, the Listener would accept, and $P(L)$ is the set of profiles from which the Speaker can be `guided' by $L$ to some state which can persuade the Listener. By definition, $T(L) \subseteq P(L)$. 


For any  set $A$ of states, $A$ is \textit{implementable} if $A = P(L)$ for some codex $L$. The set $A$ is \textit{truthfully implementable} if  $A = T(L) = P(L)$ for some codex $L$.

Let $T(\to_L) = \{s\in S\mid \forall t(s\to_L t\Rightarrow s = t)\}$. 

\begin{lemma}[\cite{glazer2012model}, Lemma 2]
\label{lemma: T of L is T of the guidance rel} For any codex $L$,
    $T(L) = T(\to_L)$.
\end{lemma}

\begin{proof}
 If $s \notin T(L)$, then $s \not\models \varphi$ for some   $\varphi\in L$; that is, $s\models a(\varphi)$ and $s\not\models z(\varphi) = \varphi_{x}$ for some $x\in \mathsf{Prop}$. Thus, $s \to_L s'$ where $s'$ is the profile that differs from $s$ only in the truth value of the variable $x$, which shows that $s \notin T(\to_L)$.
Conversely, let $s \notin T(\to_L)$, i.e.~$s \to_L t$ for some profile $t \neq s$. By definition of $\to_L$, this implies that $s\not\models\varphi$ for some   $\varphi\in L$; therefore, $s \notin T(L)$.
\end{proof}


\begin{definition}[Canonical and quasi-canonical codices]
\label{def: canonical codex}
A codex $L$ is {\em canonical} if:
\begin{enumerate}
    \item $L$ consists of complete rules.
    \item For every $s$, there is at most one $t \neq s$ such that $s \to_L t$.
    \item For every $s \in P(L) \setminus T(L)$, there is some $r \in S \setminus P(L)$ such that $r \to_L s$.
\end{enumerate}
A codex $L$ is {\em quasi-canonical} if it satisfies the conditions in the first two items above.\footnote{While the terminology `quasi-canonical' has not been introduced in \cite{glazer2012model}, it will be useful to isolate this notion for the purposes of the present paper, so we find it convenient to mention it here, rather than later on in the paper.}
\end{definition}
\begin{proposition}[ \cite{glazer2012model}, Proposition 1]
\label{prop: truthfully implementable implies canonical}
 If $A\subseteq S$ is truthfully implementable, then it is
truthfully implementable by a canonical codex.
\end{proposition}
We conclude this subsection by reporting on some examples discussed in \cite{glazer2012model}.
Let $\mathsf{Prop} = \{v_1, v_2, v_3\}$. The codex \[L = \{v_1\wedge v_2\to v_3, v_2\wedge \neg v_3\to v_1, \neg v_1\wedge \neg v_2\to \neg v_3, \neg v_1\wedge v_2\to \neg v_3, \neg v_2\wedge v_3\to \neg v_1, v_1\wedge v_3\to \neg v_2\}\]
(cf.~\cite[Section VII]{glazer2012model})  is  canonical and s.t.~$P(L) = \{(0, 0, 0), (1, 0, 0), (0, 0, 1)\}$, and $T(L) = \{(0, 0, 0), (1, 0, 0)\}$. Below, a diagram of the non reflexive edges of its corresponding guidance relation $\to_L$; in the picture, $T(L)$ is shaded in blue and $P(L)$ in red.

\begin{center}
\begin{tikzpicture}[scale=0.8, every node/.style={draw, circle, inner sep=1pt}]

\node (000) at (0, 0) {000};
\node (100) at (-2, 2) {100};
\node (010) at (0, 2) {010};
\node (001) at (2, 2) {001};
\node (110) at (-2, 4) {110};
\node (101) at (0, 4) {101};
\node (011) at (2, 4) {011};
\node (111) at (0, 6) {111};

\draw[->] (010) -- (110);
\draw[->] (001) -- (000);
\draw [->] (110) -- (111);
\draw[->] (101) -- (001);
\draw[->] (111) -- (101);
\draw[->] (011) -- (010);

\filldraw[thick, blue, opacity=0.2, rounded corners=10pt] 
     (0,-0.7) -- (-2.7, 2) -- (-2,2.7) -- (0.7,0) -- cycle;

\filldraw[thick, red, opacity=0.2, rounded corners=10pt] 
     (0,-0.9) -- (-2.9, 2) -- (-2,2.9) -- (0,0.9) -- (2,2.9) -- (2.9,2) -- cycle;

\end{tikzpicture}
\end{center}


The following  codices (cf.~\cite[Example 1, Section III]{glazer2012model}) have the same $T(-)$-set but different $P(-)$-sets:
\begin{center}
    \begin{tabular}{rl}
   $L_1 =$      & $\{v_1 \rightarrow v_2, \neg v_1 \rightarrow \neg v_2, v_1 \rightarrow v_3, \neg v_1 \rightarrow \neg v_3\}$ \\
   $L_2 =$      & $\{v_1 \rightarrow v_2, \neg v_1 \rightarrow \neg v_2, v_2 \rightarrow v_3, \neg v_2 \rightarrow \neg v_3\}$\\
   $L_3 =$      & $\{ \neg v_i \land v_{i+1} \rightarrow v_{i+2}, v_i \land \neg v_{i+1} \rightarrow \neg v_{i+2} \mid  i\in \mathbb{Z}_3 \}$
         
    \end{tabular}
\end{center}
Below, the diagrams of the non reflexive edges of their corresponding guidance relations:
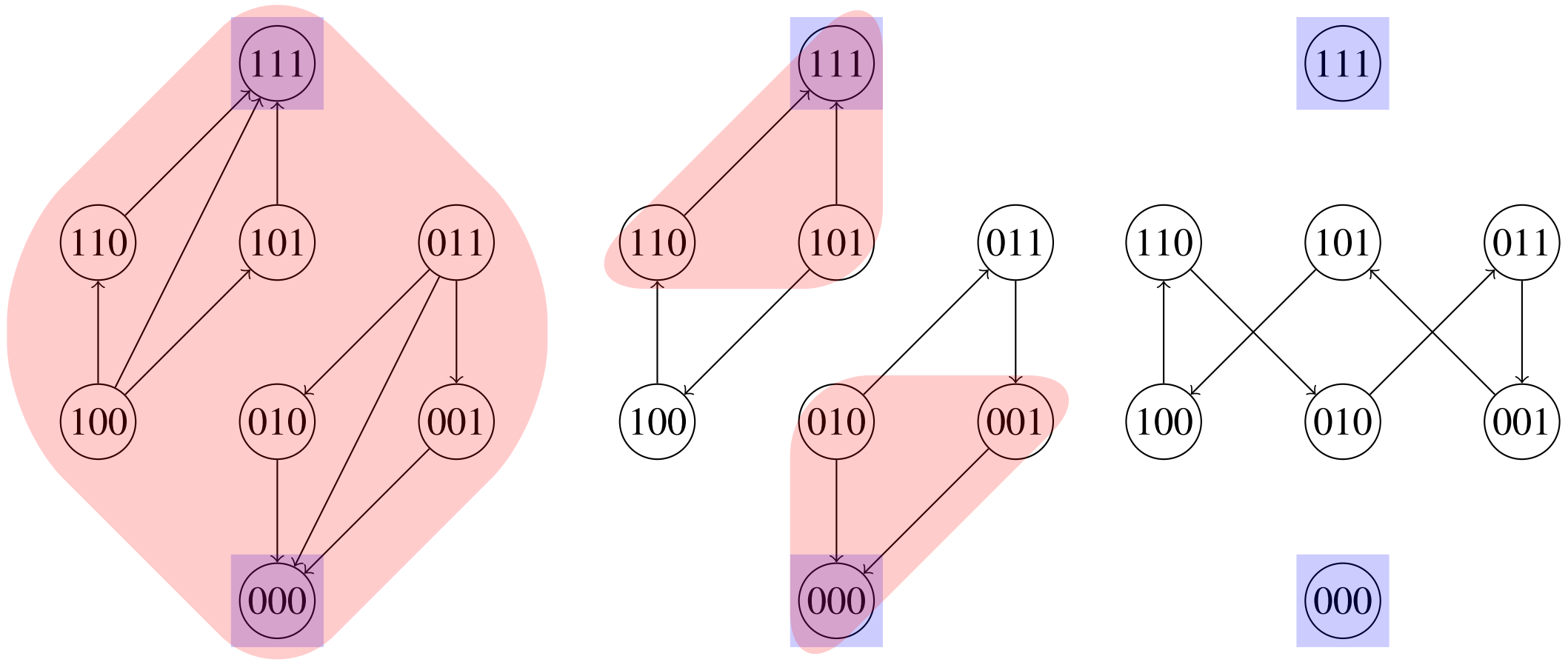
\begin{center}
\begin{tabular}{m{0.3\textwidth}m{0.3\textwidth}m{0.3\textwidth}}
  \begin{tikzpicture}[scale=0.8, every node/.style={draw, circle, inner sep=1pt}]

\node (000) at (0, 0) {000};
\node (100) at (-2, 2) {100};
\node (010) at (0, 2) {010};
\node (001) at (2, 2) {001};
\node (110) at (-2, 4) {110};
\node (101) at (0, 4) {101};
\node (011) at (2, 4) {011};
\node (111) at (0, 6) {111};

\draw[<-] (000) -- (010);
\draw[<-] (000) -- (001);
\draw[<-] (000) -- (011);
\draw[->] (100) -- (110);
\draw[->] (100) -- (101);
\draw[->] (100) -- (111);
\draw[<-] (010) -- (011);
\draw[<-] (001) -- (011);
\draw [->] (110) -- (111);
\draw[->] (101) -- (111);
\filldraw[thick, blue, opacity=0.2] (-0.5,5.5) rectangle (0.5, 6.5);

 \filldraw[thick, blue, opacity=0.2] (-0.5,-0.5) rectangle (0.5, 0.5);

  \filldraw[thick, red, opacity=0.2, rounded corners=20pt] 
    (-3, 4) -- (0, 7) -- (3, 4) --(3, 2) -- (0, -1) --
     (-3, 2) -- cycle;
\end{tikzpicture}
     &  
\begin{tikzpicture}[scale=0.8, every node/.style={draw, circle, inner sep=1pt}]

\node (000) at (0, 0) {000};
\node (100) at (-2, 2) {100};
\node (010) at (0, 2) {010};
\node (001) at (2, 2) {001};
\node (110) at (-2, 4) {110};
\node (101) at (0, 4) {101};
\node (011) at (2, 4) {011};
\node (111) at (0, 6) {111};

\draw[<-] (000) -- (010);
\draw[<-] (000) -- (001);
\draw[->] (100) -- (110);
\draw[<-] (100) -- (101);
\draw[->] (010) -- (011);
\draw[<-] (001) -- (011);
\draw [->] (110) -- (111);
\draw[->] (101) -- (111);
\filldraw[thick, blue, opacity=0.2] (-0.5,5.5) rectangle (0.5, 6.5);

 \filldraw[thick, blue, opacity=0.2] (-0.5,-0.5) rectangle (0.5, 0.5);
 \filldraw[thick, red, opacity=0.2, rounded corners=20pt] 
    (-3, 3.5) -- (0.5, 7) -- (0.5, 3.5) -- cycle;
    \filldraw[thick, red, opacity=0.2, rounded corners=20pt] 
     (-0.5, 2.5) -- (3, 2.5)  --(-0.5, -1) -- cycle;
\end{tikzpicture}     
     &    
 \begin{tikzpicture}[scale=0.8, every node/.style={draw, circle, inner sep=1pt}]

\node (000) at (0, 0) {000};
\node (100) at (-2, 2) {100};
\node (010) at (0, 2) {010};
\node (001) at (2, 2) {001};
\node (110) at (-2, 4) {110};
\node (101) at (0, 4) {101};
\node (011) at (2, 4) {011};
\node (111) at (0, 6) {111};

\draw[->] (100) -- (110);
\draw[->] (110) -- (010);
\draw[->] (010) -- (011);
\draw[->] (001) -- (101);
\draw[->] (011) -- (001);
\draw[->] (101) -- (100);
 \filldraw[thick, blue, opacity=0.2] (-0.5,5.5) rectangle (0.5, 6.5);

 \filldraw[thick, blue, opacity=0.2] (-0.5,-0.5) rectangle (0.5, 0.5);
 
\end{tikzpicture}     
     \\
     
\end{tabular}
\end{center}

The set $A = \{(1, 0, 0), (0, 1, 0), (0, 0, 1)\}$, represented below together with the non-reflexive edges of the neighbouring relation, is not implementable (cf.~\cite[Example 2, Section III]{glazer2012model}):

\begin{center}
     \begin{tikzpicture}[scale=0.8, every node/.style={draw, circle, inner sep=1pt}]

\node (000) at (0, 0) {000};
\node (100) at (-2, 2) {100};
\node (010) at (0, 2) {010};
\node (001) at (2, 2) {001};
\node (110) at (-2, 4) {110};
\node (101) at (0, 4) {101};
\node (011) at (2, 4) {011};
\node (111) at (0, 6) {111};
 \filldraw[thick, blue, opacity=0.2] (-2.5,1.5) rectangle (2.5, 2.5);
\draw (000) -- (100);
\draw[-] (000) -- (010);
\draw[-] (000) -- (001);
\draw[-] (100) -- (101);
\draw[-] (100) -- (110);
\draw (010) -- (110);
\draw[-] (010) -- (011);
\draw (001) -- (101);
\draw[-] (001) -- (011);
\draw [-] (110) -- (111);
\draw[-] (101) -- (111);
\draw (011) -- (111);

\end{tikzpicture}  
\end{center}
Indeed, assuming for contradiction that $A = P(L)$ for some codex $L$,

\begin{itemize}
\item \textbf{Case 1}: if $T(L) = P(L) = A$, i.e.~$A$ is truthfully implementable, then by  Proposition  \ref{prop: truthfully implementable implies canonical} we can assume w.l.o.g.~that $L$ is canonical, and since $s_0 = (0, 0, 0)\notin A =  T(L)$,  there is a complete rule $\varphi\in L$ s.t.~$s_0\not\models \varphi$; this violation implies that $s_0\to_L s$ for some $s \neq s_0$ and, since the rules in $L$ are complete and hence $\to_L$ is contained in the neighbourhood  relation $N$, $s\in A$. Hence, $s_0\in P(L) = A$, which  contradicts $s_0\notin A$. 

\item \textbf{Case 2}: if $T(L)\subsetneq A = P(L)$, then  one of the profiles in $A = P(L)$, w.l.o.g.~$s = (0, 0, 1)$, is not in $T(L)$. Since $(0, 0, 1)\in A = P(L)$, there must be another profile in $A$, w.l.o.g. $s' = (0, 1, 0)$, such that $(0, 0, 1) \rightarrow_L (0, 1, 0)$. This requires that $\neg v_1 \rightarrow v_2$ (and $\neg v_1\rightarrow \neg v_3$) be in the codex; indeed, $v_1$ is the only variable on which $s$ and $s'$ agree.  However, in that case, $(0, 0, 0) \rightarrow_L (0, 1, 0) \in T(L)$ and therefore $(0, 0, 0) \in P(L) = A$ although $(0, 0, 0) \not\in A$, a contradiction.

\end{itemize}

Although the above set $A$ is not implementable, its complement is. Let $A^c = S \setminus A$. Consider the codex $L^c$ that consists of the three rules $v_i \rightarrow v_j$, where $j \neq i+1$; i.e.:
\[L^c = \{v_1\to v_3, v_2\to v_1, v_3\to v_2\}\]
The picture below represents the diagram of the non-reflexive edges of the guidance relation $\to_L$; the sets $T(L^c)$ and $P(L^c)$ are  shaded in blue and red, respectively.
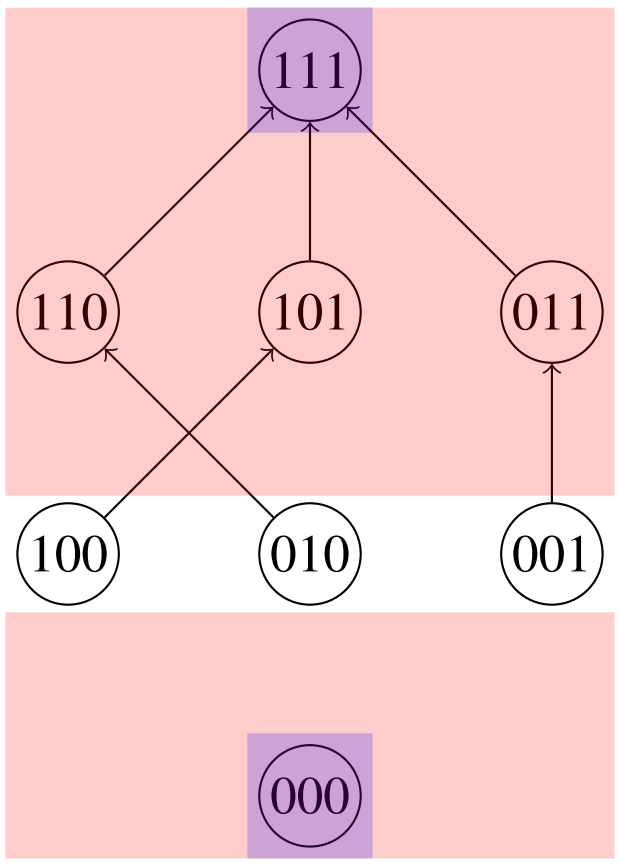
\begin{center}
\begin{tikzpicture}[scale=0.8, every node/.style={draw, circle, inner sep=1pt}]

\node (000) at (0, 0) {000};
\node (100) at (-2, 2) {100};
\node (010) at (0, 2) {010};
\node (001) at (2, 2) {001};
\node (110) at (-2, 4) {110};
\node (101) at (0, 4) {101};
\node (011) at (2, 4) {011};
\node (111) at (0, 6) {111};

\draw[->] (101) -- (111);
\draw[->] (010) -- (110);
\draw[->] (001) -- (011);
\draw [->] (110) -- (111);
\draw[->] (100) -- (101);
\draw [->] (011) -- (111);
 \filldraw[thick, red, opacity=0.2] (-2.5,2.5) rectangle (2.5, 6.5);

 \filldraw[thick, red, opacity=0.2] (-2.5,-0.5) rectangle (2.5, 1.5);
 \filldraw[thick, blue, opacity=0.2] (-0.5,5.5) rectangle (0.5, 6.5);

 \filldraw[thick, blue, opacity=0.2] (-0.5,-0.5) rectangle (0.5, 0.5);
\end{tikzpicture}
\end{center}
\subsection{Basic formalism in `Magical implementation'}
\label{ssec: magical}
 In \cite{glazer2024magical}, persuasion situations are analyzed in which  (at least) two Speakers   are in a dispute over a single indivisible prize, which the Listener will assign. The set $S$ (usually $S = 2^K$ with $K\geq 3$)   is partitioned into   subsets  $W_i$ 
 where $W_i = \{s\in S\mid i $ is assigned the prize if $i$ declares $s\}$ denotes the {\em winning set} of Speaker $i$. 
 \begin{definition}
 \label{def:implementation problem}
     An  {\em implementation problem} is a tuple $(S, \{W_i\mid i\in I\})$ such that $S$ is a set  and $\{W_i\mid i\in I\}$ is a partition of $ S$. 
 \end{definition}
 Since  $\{W_i\mid i\in I\}$ is a partition of $ S$, the Listener can assume that exactly one Speaker is truthful, namely the unique $i$ such that the {\em true state} $t\in W_i$. To identify which Speaker is truthful, 
the Listener  constructs a codex $L_i$ for each Speaker $i$; this codex encodes   the circumstances in which $i$  deserves to receive the prize. 
Speaker $i$ declares 
 a state   $s\in  T(L_i)$, truthfully or not (hence, $P(L_i) = S$). It is assumed that, if $i$'s declaration is not truthful, the state declared by $i$  is `suggested' by $\to_{L_i}$, i.e.~$i$ declares some $s\in S$ s.t.~$t\to_{L_i} s$. 
 \begin{definition}
  \label{def:magical implementation}   

 The  codices $\{L_i\mid i\in I\}$ {\em magically implement} $(S, \{W_i\mid i\in I\})$ if:
 \begin{enumerate}
     \item $T(L_i) = W_i$ for each $i$; 
     \item For all $s, t \in S$, if $s \to_{L_i} t$ 
     then $t \not\to_{L_j} s$ for all $j\neq i$.
 \end{enumerate}
 \end{definition}

 \begin{remark}
     The second item of the previous definition allows the Listener to apply what the authors of \cite{glazer2024magical} refer to as the {\em one-way cheating principle}. When $P(L_i)\cup\{s\in S\ \mid\ \forall t\in S (s\rightarrow_{L_i} t\Rightarrow s=t)\} = S$, the one-way cheating principle implies  that the Listener can infer the true state from the statements of all Speakers. In the following sections, this will always be assumed (cf.\ Definition \ref{def:TP:rel}, condition R3 and its strengthenings). However, if this is not the case, then, in principle, the Listener is not always able to discern the true state, even if the codices magically implement the partition. For instance, in the example below, if, in $L_{Con}$, the rule $\lnot v_2\land\lnot v_3\rightarrow \lnot v_1$ is replaced by $v_1\land \lnot v_3\rightarrow v_2$, then, when the true state is $(1,0,0)$, the Listener cannot infer this, even though the codices magically implement the partition.  
 \end{remark}
To illustrate the application of the one-way cheating principle in action, in \cite{glazer2024magical}, the implementation problem $(S, W_{Pro}, W_{Con})$ is discussed, in which $S = 2^{\mathsf{Prop}}$ for $\mathsf{Prop} = \{v_1, v_2, v_3\}$, and  $W_{Pro} = \{(1, 0, 0), (0, 1, 0), (0, 0, 1)\}$\footnote{In what follows, we will abbreviate e.g.~$(1, 0, 0)$ as $100$, and so on.} and $W_{Con} = S\setminus W_{Pro}$. It is easy to verify that the codices in the table below magically implement $(S, W_{Pro}, W_{Con})$. 
\begin{table}[h!]
\centering
\begin{tabular}{|l|l|}
\hline
\textbf{$L_{Pro}$} & \textbf{$L_{Con}$} \\
\hline
\(\neg v_1 \land v_2 \to \neg v_3\) & \(v_2 \land \neg v_3 \to v_1\) \\
\(v_1 \land v_2 \to v_3\) & \(\neg v_1 \land v_3 \to v_2\) \\
\(v_1 \land v_3 \to \neg v_2\) & \(\neg v_2 \land \neg v_3 \to \neg v_1\) \\
\(v_1 \land \neg v_2 \to \neg v_3\) & \\
\(\neg v_1 \land \neg v_2 \to v_3\) & \\
\hline
\end{tabular}
\end{table}

Notice that each codex is quasi-canonical but not canonical (cf.~Definition \ref{def: canonical codex}). The following diagrams represent (the non reflexive edges of) the  guidance relation associated with each codex; $W_{Pro}$ is shaded in red and $W_{Con}$ in blue.
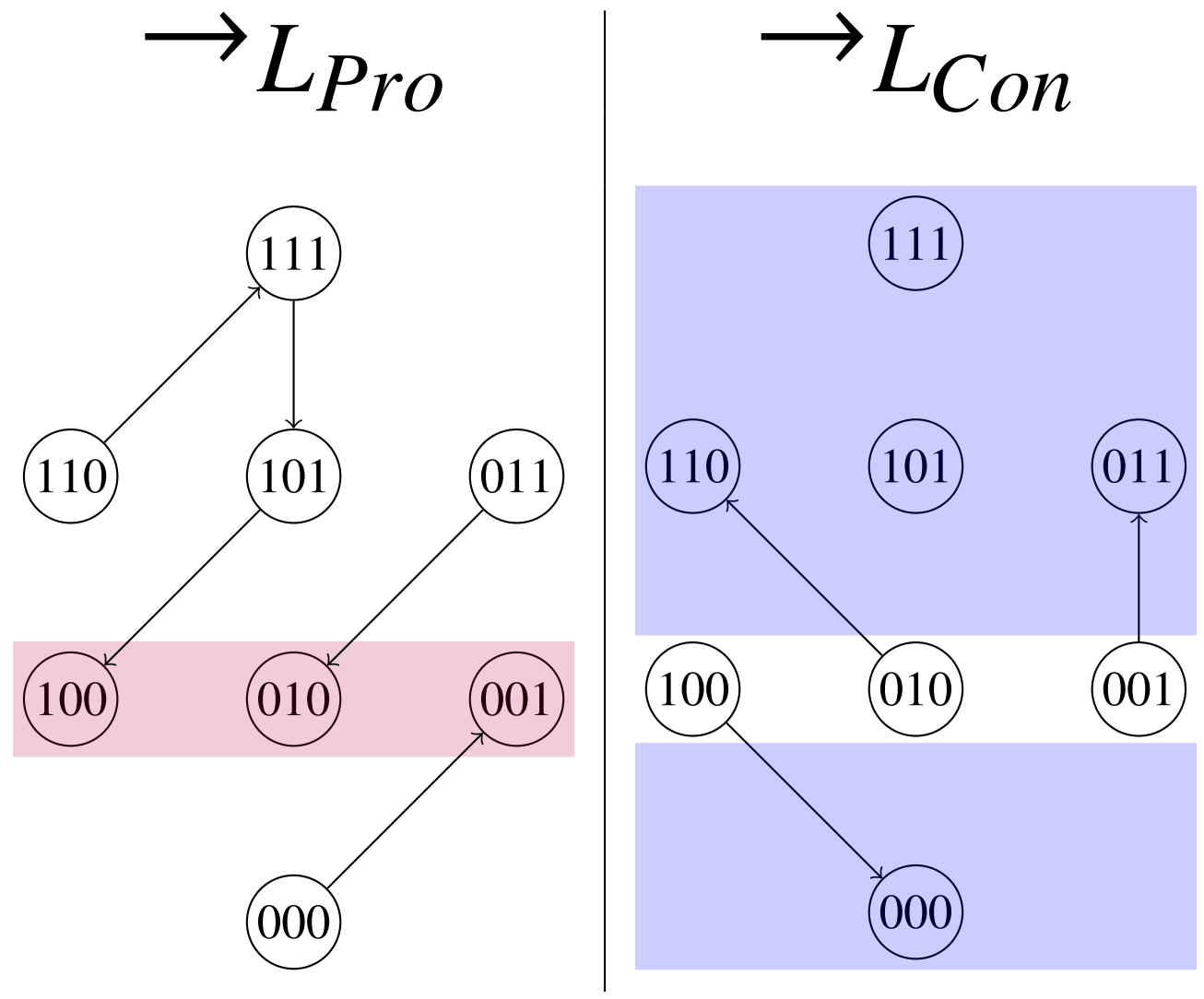
\begin{center}
    \begin{tabular}{c|c}
    \Huge{$\to_{L_{Pro}}$} & \Huge{$\to_{L_{Con}}$}\\
    &\\
   \begin{tikzpicture}[scale=0.8, every node/.style={draw, circle, inner sep=1pt}]

\node (000) at (0, 0) {000};
\node (100) at (-2, 2) {100};
\node (010) at (0, 2) {010};
\node (001) at (2, 2) {001};
\node (110) at (-2, 4) {110};
\node (101) at (0, 4) {101};
\node (011) at (2, 4) {011};
\node (111) at (0, 6) {111};
 \filldraw[thick, purple, opacity=0.2] (-2.5,1.5) rectangle (2.5, 2.5);
\draw[->] (000) -- (001);
\draw[<-] (100) -- (101);
\draw[<-] (010) -- (011);
\draw [->] (110) -- (111); 
\draw[<-] (101) -- (111);

\end{tikzpicture}      & 
\begin{tikzpicture}[scale=0.8, every node/.style={draw, circle, inner sep=1pt}]

\node (000) at (0, 0) {000};
\node (100) at (-2, 2) {100};
\node (010) at (0, 2) {010};
\node (001) at (2, 2) {001};
\node (110) at (-2, 4) {110};
\node (101) at (0, 4) {101};
\node (011) at (2, 4) {011};
\node (111) at (0, 6) {111};

\draw[<-] (000) -- (100);
\draw[->] (010) -- (110);
\draw[->] (001) -- (011);
 \filldraw[thick, blue, opacity=0.2] (-2.5,2.5) rectangle (2.5, 6.5);

 \filldraw[thick, blue, opacity=0.2] (-2.5,-0.5) rectangle (2.5, 1.5);
 
\end{tikzpicture}
\\
     
    \end{tabular}
\end{center}

It is assumed that the Listener knows the cheating procedure applied by each Speaker $i$. If the true state $t\notin W_i$, then $i$ 
will search for a state $s\in T(L_i)$  by advancing along a path which starts at $t$ following the guidance relation $\to_{L_i}$. 
%
For any Speaker $i$, each rule in $L_i$  has two effects: it excludes one state in $S\setminus W_i$ and directs  Speaker $i$ to a different (unique) state. For example, the rule \(v_1 \land \neg v_2 \to \neg v_3\) on $Pro$’s list is violated by \(101\), and if \(101\) happens to be the true state, it guides $Pro$ to declare \(100\).

To see the one-way cheating principle at work, assume, for example, that the true state is  $t\in \{111, 110, 101 \}\subseteq W_{Con}$. Since $t\in W_{Con}$, agent
 $Con$ will truthfully declare $t$, while, being iteratively guided by $\to_{L_{Pro}}$, agent $Pro$  declares $100$. 
Hence, upon receiving e.g.~$110$ from   $Con$ and $100$ from $Pro$, the Listener concludes that $Con$ is the truthful one, since $100$ is the state $Pro$ would declare in case the true state was $110$, while, if $100$ were the true state,  $Con$ would have declared $000$ instead of $110$.

Another example of implementation problem mentioned in \cite[Section 3]{glazer2024magical} is $(S, W_Y, W_Z)$ with $S = W_Y\cup W_Z$ and $W_Z = \{z_1, z_2, z_3\}$ and $W_Y = \{y_1,\ldots, y_5\}$. Interestingly, this example does not involve codices, and demonstrates that problems such as $(S, W_Y, W_Z)$ can still be {\em magically implemented by pairs $(\to_Y, \to_Z)$ of  binary  relations on $S$}, provided these relations satisfy conditions analogous to those  of Definition \ref{def:magical implementation}, i.e.:\footnote{In Section \ref{sec:TP relations and game-theoretic},  the notion of magical implementation via relations is introduced more  systematically   (cf.~Definition \ref{def: magical implementation extended set-theoretic}).}
\begin{enumerate}
    \item 
$T(\to_U): =  \{s\in S\mid \forall t(s\to_U t\Rightarrow s = t)\} = W_U$ for any $U\in \{Y, Z\}$, and 
\item for all $U, V\in \{Y, Z\}$ s.t.~$U\neq V$  and  all $s, t\in S$, $s\to_U t$ implies $t\not\to_V s$.
\end{enumerate} In the case of this example, it can  readily be verified by inspection \label{page:magical implementation via relations in the example} that the relations $\to_{Y}$ and $\to_Z$ represented below do satisfy these conditions.\footnote{\label{footnote: relationship between magical implementation two ways} Hence,  Definition \ref{def:magical implementation} can equivalently be formulated by first defining the notion of magical implementation via relations as illustrated above and then by stipulating that a set of codices magically implement an implementation problem if so do their associated guidance relations.}  Hence, if $Y$ and $Z$ are guided by these relations, the Listener can establish who the truthful Speaker is whichever state in $S$ is the true one.  For instance, if $Y$ declares $z_1$ and $Z$ declares $y_2$, then the Listener concludes that $Y$ is the truthful one, since $y_2$ is the state $Z$ would declare if $t = z_1$, while if $t = y_2$, then $Y$ would have declared $z_2$, and similarly for the remaining situations.

The one-way cheating principle can be understood as a particular instance of the Tiglath-Pileser principle discussed in the introduction.
Indeed, the latter principle provides a general epistemic criterion for evaluating untrustworthy declarations by reconstructing the set of states that could have generated them under constrained distortion.
The one-way cheating principle adapts and applies this criterion to a multi-agent setting in which the distortion mechanisms (guidance relations) are sufficiently structured so as to allow, under favourable circumstances,  to  uniquely identify the true state and the truthful speaker.
In Section \ref{ssec:Magical implementation via relations}, we will suitably generalize the magical implementation framework to formally account for a wider-scoped applicability of the Tiglath-Pileser principle. 

\begin{center}
    \begin{tabular}{c|c}
    \Huge{$\to_{Y}$} & \Huge{$\to_{Z}$}\\
    &\\

\begin{tikzpicture}[scale=1.2]  

\node (z1) at (0,3) {};
\node (z2) at (0,2) {};
\node (z3) at (0,1) {};
\node (y1) at (2,3) {};
\node (y2) at (2,2) {};
\node (y3) at (2,1) {};
\node (y4) at (2,0) {};
\node (y5) at (2,-1) {};

\draw[<-] (z1) -- (y1);
\draw[<-] (z2) -- (y2);
\draw[<-] (z3) -- (y3);
\draw[<-] (z1) -- (y4);
\draw[<-] (z2) -- (y5);

\node[anchor=east] at (z1) {$z_1$};
\node[anchor=east] at (z2) {$z_2$};
\node[anchor=east] at (z3) {$z_3$};
\node[anchor=west] at (y1) {$y_1$};
\node[anchor=west] at (y2) {$y_2$};
\node[anchor=west] at (y3) {$y_3$};
\node[anchor=west] at (y4) {$y_4$};
\node[anchor=west] at (y5) {$y_5$};


 \filldraw[thick, purple, opacity=0.2] (-0.5,0) rectangle (0.5, 3.5);
\end{tikzpicture}      & 
\begin{tikzpicture}[scale=1.2]  

\node (z1r) at (4,3) {};
\node (z2r) at (4,2) {};
\node (z3r) at (4,1) {};
\node (y1r) at (6,3) {};
\node (y2r) at (6,2) {};
\node (y3r) at (6,1) {};
\node (y4r) at (6,0) {};
\node (y5r) at (6,-1) {};

\draw (z1r)[->] -- (y2r);
\draw (z2r)[->] -- (y3r);
\draw (z3r)[->] -- (y4r);

\node[anchor=east] at (z1r) {$z_1$};
\node[anchor=east] at (z2r) {$z_2$};
\node[anchor=east] at (z3r) {$z_3$};
\node[anchor=west] at (y1r) {$y_1$};
\node[anchor=west] at (y2r) {$y_2$};
\node[anchor=west] at (y3r) {$y_3$};
\node[anchor=west] at (y4r) {$y_4$};
\node[anchor=west] at (y5r) {$y_5$};

\filldraw[thick, blue, opacity=0.2] (5.5,-1.3) rectangle (6.5, 3.5);
 
\end{tikzpicture}
\\
     
    \end{tabular}
\end{center}

\section{More on codices and guidance relations}\label{Sec:Tiglath-Pileser:Relations}

In the present section, we further investigate the two-way relationship between codices and (guidance) relations. Specifically,  in Section \ref{ssec:codices from relations} we intrinsically characterize the class of  guidance relations associated with quasi-canonical codices; in Section \ref{ssec:separated codices}, we introduce and study a class of codices which will be relevant to the analysis of Section \ref{Sec:Dynamically:Formed:Relations}; in Section \ref{ssec: from relations to codices}, we discuss a construction which allows us to recast a problem magically implementable via relations (cf.~discussion at the end of Section \ref{ssec: magical}) as a problem magically implementable via codices. 

\subsection{Quasi-canonical codices and guiding relations }  
\label{ssec:codices from relations}

\begin{definition}
\label{def:pre-guiding relation}
 Let  $S = 2^K$. A reflexive and antisymmetric binary relation ${\to}$ on $S$ is a {\em guiding relation} if it satisfies the following conditions:
\begin{enumerate}
    \item ${\to}\subseteq N$;

\item for every $s\in S$, there is at most one $s'\in S$ s.t.~$s\neq s'$ and $s\to s'$. \end{enumerate}
We let $T(\to): = \{s\in S\mid \forall t(s\to t\Rightarrow s = t)\}$, and for any $s\in S\setminus T(\to)$, we let $s_\to\in S$ denote the unique profile $s'$ satisfying item 2 above.
\end{definition}
As we will see next, the notion of guid{\em ing} relation is a characterization of the guid{\em ance} relations associated with quasi-canonical codices, formulated without any reference to codices. Examples of guiding relations occur in every magical implementation scenario discussed in Section \ref{ssec: magical}. 
\begin{lemma}
\label{lemma: L quasi canonical implies to-L guiding}
  If $L$ is a quasi-canonical codex, then $\to_L$ is a guiding relation. 
\end{lemma}
\begin{proof}
    By construction, $\to_L$ is reflexive; item 1 of  Definition \ref{def:pre-guiding relation}  follows from the rules of $L$ being complete; item 2 immediately follows from   item 2 of the definition of pre-canonical codex. To complete the proof, let us show contrapositively that the coherence of $L$ implies that $\to_L$ is antisymmetric. Let $s, t\in S$ s.t.~$s\neq t$ and $s\to_L t$ and $t\to_L s$; then,  $s, t\notin T(L)$ and $sNt$; hence, some $\varphi, \psi\in L$ exist s.t.
    \[s\models a(\varphi),\; t\models a(\varphi),\; t\models z(\varphi),\;s\not\models z(\varphi)\]\[t\models a(\psi),\;s\models a(\psi),\;s\models z(\psi),\;t\not\models z(\psi).\]
    Hence, $\varphi$ and $\psi$ are s.t.~$a(\varphi)$ and $a(\psi)$ are jointly satisfiable, and since they are complete rules, from the conditions above it also follows that $z(\varphi)$ and $z(\psi)$ share the one variable occurring in them, and are hence not jointly satisfiable, which proves that $L$ is not coherent.
\end{proof}
As discussed above, Definition \ref{def:pre-guiding relation} implies that $s_\to$ exists and $sNs_{\to}$ for any $s\in S\setminus T(\to)$, and hence $\varphi(s, s_{\to})$ (see \eqref{eq:phi-s-s-prime}) is well defined.  Let  \begin{equation}
     L_{\to}: = \{\varphi(s, s_\to)\mid s\in S\setminus T(\to)\}.
 \end{equation}
\begin{lemma}
  If $\to$ is a guiding relation, then $L_\to$ is a quasi-canonical codex. 
\end{lemma}
\begin{proof}

By construction,  $L_{\to}$ consists of complete rules, and for every $s\in S$ there is at most one $t\neq s$ such that $s\to_{L_\to} t$, which shows that $L_\to$ is pre-canonical. To finish the proof, arguing contrapositively, let us assume  that $L_\to$ is not coherent, and let $\varphi, \psi\in L_\to$ s.t.~$a(\varphi)$ and $ a(\psi)$ are jointly satisfiable while $z(\varphi)$ and $ z(\psi)$ are not. Let $s, s'\in S\setminus T(\to)$ s.t.~$\varphi = \varphi(s, s_\to)$
and $\psi = \varphi(s', s'_\to)$, with \[s\models a(\varphi),\; s_\to\models a(\varphi),\; s_{\to}\models z(\varphi),\;s\not\models z(\varphi)\]\[s'\models a(\psi),\;s'_{\to}\models a(\psi),\;s'_{\to}\models z(\psi),\;s '\not\models z(\psi).\] Since $\varphi$ and $\psi$ are complete rules, $a(\varphi)$  and $a(\psi)$ each contain $K-1$ variables; since $z(\varphi)$ and $ z(\psi)$ are not jointly satisfiable, it follows w.l.o.g.~that $z(\varphi) = v$ and $z(\psi) = \neg v$ for some $v\in \mathsf{Prop}$, hence $a(\varphi)$  and $a(\psi)$ have exactly the same variables, and hence the assumption that they are jointly satisfiable  implies that $a(\varphi) = a(\psi)$. Moreover, $s\not\models z(\varphi)$ iff $s\models z(\psi)$ and $s'\not\models z(\psi)$ iff $s'\models z(\varphi)$, which imply that $s = s'_{\to}$ and $s' = s_{\to}$, and hence $s\to s'$ and $s'\to s$ and $s\neq s'$, showing that $\to$ is not antisymmetric. 
\end{proof}
Recall that $T(L) = T(\to_L)$  for any codex $L$  (cf.~Lemma \ref{lemma: T of L is T of the guidance rel}). The next proposition embeds this identity into a perfect correspondence between  quasi-canonical codices and guiding relations.
 \begin{proposition}
 For any guiding relation $\to$ and any quasi-canonical codex $L$,
 \begin{enumerate}
     \item 
    $T(\to) = T(L_\to) $ and ${\to_{L_\to}} = {\to}$;
     \item    $L = L_{\to_L}$.
      \end{enumerate}
 \end{proposition}
\begin{proof}
1. By construction, $s\not\models \varphi(s, s_\to)$ for every $s\in S\setminus T(\to)$, which shows contrapositively that $T(L_\to)\subseteq T(\to)$. Conversely, if $t\in T(\to)$, then $t\neq s$ for every $s\in S\setminus T(\to)$, and so, fixing one such $s$, $s(v)\neq t(v)$ for at least one $v\in V$. 
Then either $t\not \models a(\varphi(s, s_\to))$, and hence $t\models \varphi(s, s_\to)$, or $t \models a(\varphi(s, s_\to))$, and hence $t = s_\to\models \varphi(s, s_\to)$. This proves the first identity. As to the second identity, since the two relations are reflexive, it is enough to show that if $s, t\in S$ and $s\neq t$, then $s\to_{L_\to} t$ iff $s\to t$. Let us assume that $s\to_{L_\to} t$; since $s\neq t$, to show that $s\to t$, it is enough to show that $t = s_\to$.  Let   $v\in V$ s.t.~$s(v) \neq t(v)$, and let $\varphi$ be the witness rule for $v$, i.e.~$s\models a(\varphi)$, $t\models a(\varphi)$ and $t\models z(\varphi)\in \{v, \neg v\}$. Hence, $s(v) \neq t(v)$ implies that $s\not\models \varphi$, and since, by construction, each  rule in $ L_{\to}$ is of the form $\varphi(s', s'_\to)$ for some $s'\in S\setminus T(\to)$ and can only be falsified at $s'$, it follows that $\varphi = \varphi(s, s_\to)$. Moreover,  since $\varphi(s, s_\to)$ is complete and $s\models a(\varphi)$, and $t\models a(\varphi)$, then $v$ is the only variable on which $s$ and $t$ disagree, and moreover, $t(v) = s_{\to}(v)$, which implies that $t = s_\to$, as required. Conversely, if $s\to t$ and $s\neq t$, then $s\in S\setminus T(\to)$ and $t = s_\to$, hence the rule $\varphi(s, s_\to)$  witnesses that $s\to_{L_{\to}} t$ for the unique variable on which $s$ and $t$ differ.

2. By definition and Lemma \ref{lemma: T of L is T of the guidance rel}, $L_{\to_{L}} = \{\varphi(s, s_{\to_L})\mid s\in S\setminus T(\to_L)\} = \{\varphi(s, s_{\to_L})\mid s\in S\setminus T(L)\}$.

Notice preliminarily that any complete rule $\varphi$ coincides with the rule $\varphi(s, s')$ (see \eqref{eq:phi-s-s-prime}) for exactly two states $s, s'\in S$ s.t.~$s\neq s'$ and $sNs'$.

 Let $\varphi\in L$; the assumption that $L$ is quasi-canonical implies that $\varphi$ is a complete rule; therefore, let $s, s'\in S$ be the unique states s.t.~$s\neq s'$, $sNs'$, and $\varphi = \varphi(s, s')$. Then $s\models a(\varphi)$, $s'\models a(\varphi)$, $s'\models z(\varphi)$ and $s\not\models z(\varphi)$, which shows that  $s\in S\setminus T(L)$,  and  $s' = s_{\to_L}$, completing the proof  that $\varphi\in L_{\to_L}$. For the converse inclusion, let $s\in  S\setminus T(\to_L) = S\setminus T(L)$; hence, $s\not \models \varphi$ for some complete rule $\varphi\in L$ which, by the preliminary observation, coincides with $\varphi(s, s')$ for a unique state $s'\in S$ s.t.~$s\neq s'$ and $sNs'$. Hence, $s' = s_{\to_L}$. 
\end{proof}

Although the codices mentioned in the definition of  magical implementation (cf.~Definition \ref{def:magical implementation}) are not required to be  quasi-canonical  in \cite{glazer2024magical}, the defining  properties of quasi-canonical codices are key to the identification of the truthful Speaker via the application of the one-way cheating principle in all the concrete  examples discussed in \cite{glazer2024magical}. Furthermore, as mentioned when discussing  the last example of Section \ref{ssec: magical} (see page \pageref{page:magical implementation via relations in the example}), also in the context of the implementation problem $(S, W_Y, W_Z)$, in which the one-way cheating principle is discussed without reference to codices, the relations which allow for the applicability of the one-way cheating principle satisfy  the defining properties\footnote{More specifically, all the conditions which are applicable are satisfied; of course, item 1 of Definition \ref{def:pre-guiding relation}, requiring guiding relations to be subsets of the neighbouring relation $N$ is not applicable if $S$ is not a set of profiles.} of {\em guiding relations} which, as we have just seen,  
provide an intrinsic characterization of the guidance relations associated with {\em quasi-canonical} codices. In Section \ref{ssec:Magical implementation via relations}, we  generalize the magical implementation framework  to  relations on {\em arbitrary} sets $S$ (that is, beyond the context of sets $S = 2^K$ of profiles).

\subsection{Separated codices and their guidance relations}
\label{ssec:separated codices}

In the present section, we introduce a subclass of codices which will become useful in the developments of Section \ref{Sec:Dynamically:Formed:Relations}.

\begin{definition}[Separated codex]
A codex $L$ is {\em separated} if two disjoint subsets $A, Z\subseteq \mathsf{Prop}$ exist such that $Var(a(\varphi))\subseteq A$ and $Var(z(\varphi))\subseteq Z$ for all $\varphi\in L$.    
\end{definition}

The subclass of separated codices  will play an important role in the analysis of Section \ref{Sec:Dynamically:Formed:Relations}, where a distinction between verifiable and non-verifiable variables is made in order to capture the fact that the speaker is constrained in their statements by the (partial) knowledge of the listener.

\begin{proposition}
    If $L$ is a separated codex, then $\to_L$ is transitive.
\end{proposition}
\begin{proof} Notice preliminarily that, by definition of $\to_L$, if $s, s'\in S = 2^K$ and $s\to_L s'$ then  $s$ and $s'$ can only differ on variables occurring in consequents of rules of $L$, and since $L$ is separated, the variables on which $s$ and $s'$ differ cannot occur in antecedents of rules, hence $s$ and $s'$  have to agree on $a(\phi)$ for all $\phi\in L$. 
    Let $s, t, u\in S$ such that $s\to_L t$ and $t\to_L u$.  To show that $s\to_L u$, let $v\in \mathsf{Prop}$ such that $s(v)\neq u(v)$. Then  either $s(v)\neq t(v) = u(v)$ or $s(v) = t(v) \neq u(v)$. If $s(v)\neq t(v) = u(v)$, then the assumption that $s\to_L t$ implies that some $\varphi\in L$ exists such that $z(\varphi)\in \{v, \neg v\}$, and  $s\models a(\varphi)$, $t\models a(\varphi)$, $s\not \models z(\varphi)$, $t\models z(\varphi)$. As noted preliminarily, $t\to_L u$ implies that $u\models a(\varphi)$. Since $t(v) = u(v)$, it follows that $u\models z(\varphi)$, as required. 

    If $s(v) = t(v) \neq u(v)$, then the assumption that $t\to_L u$ implies that some $\varphi\in L$ exists such that $z(\varphi)\in \{v, \neg v\}$, and  $t\models a(\varphi)$, $u\models a(\varphi)$, $t\not\models z(\varphi)$, $u\models z(\varphi)$. As noted preliminarily, $s\to_L t$ implies that $s\models a(\varphi)$. Since $s(v) = t(v)$, it follows that $s\not\models z(\varphi)$, as required.     
\end{proof}

\subsection{From relations to codices}
\label{ssec: from relations to codices}
At the end of Section \ref{ssec: magical}, we discussed  a problem $(S, W_Y, W_Z)$  which is magically implementable via relations rather than via codices. In this context, we also noticed (cf.~Footnote \ref{footnote: relationship between magical implementation two ways}) that the definition of magical implementation of \cite{glazer2024magical} (cf.~Definition \ref{def:magical implementation}) can equivalently be restated purely in terms of relations rather than in terms of codices.  In this subsection, we illustrate, in the context of this same example,  how $(S, W_Y, W_Z)$  can be recast as a problem $(S^P, W^P_Y, W^P_Z)$ implementable by means of codices, simply by letting the set $S =\{y_1,\ldots, y_5, z_1, z_2, z_3\}$, the domain of the original problem, be the set $\mathsf{Prop}$  of proposition {\em variables}. Under this stipulation, the  relations $\to_Y$ and $\to_Z$ represented in the last picture of Section \ref{ssec: magical} give rise to  the following 
codices: 
\[L_Y: = \{\neg y_j\mid 1\leq j\leq 5\}\cup \{ y_j\to z_i\mid y_j\to_Y z_i\}\quad\quad L_Z: =\{\neg z_j\mid 1\leq j\leq 3\}\cup \{ z_j\to y_i\mid z_j\to_Z y_i\}. \]
Let $S^P: = 2^{\mathsf{Prop}}\cong 2^8$ be the set of profiles. We assume w.l.o.g.~that each $s\in S^P$ is of the form $\overline{y}_s\overline{z}_s$, where $\overline{y}_s\in 2^5$ (resp.~$\overline{z}_s\in 2^3$) is the tuple corresponding to the truth values of the proposition variables $y_1,\ldots, y_5$ (resp.~$z_1,\ldots, z_3$). 
Then \[W^P_Y := T(L_Y) = \{\overline{0}\overline{z}  \mid \overline{z}\in 2^3\}\  \text{ and }\  W^P_Z: = T(L_Z) = \{\overline{y}\overline{0}  \mid \overline{y}\in 2^5\}.\]
Notice that $\overline{0}\in W^P_Y\cap W^P_Z\neq \varnothing$ and $\overline{1}\notin W_Y\cup W_Z\neq S$, hence  $(S^P, W^P_Y, W^P_Z)$ is not an implementation problem as per Definition \ref{def:implementation problem}.  We propose the following extension of the definition of magical implementation to structures $(S, \{W_i\mid i\in I\})$ s.t.~$\{W_i\mid i\in I\}$ is not necessarily a partition of $S$:\footnote{Notice that, under this relaxation, the two Speakers can declare states in the intersection of their winning sets, and hence there is no longer a guarantee that the Listener is able to reconstruct the true state from their declarations.}
\begin{definition}
    \label{def: magical implementation extended}
     The  codices $\{L_i\mid i\in I\}$ {\em magically implement} $(S, \{W_i\mid i\in I\})$ if, for all $i\in I$  
     and  $s, t \in S$, 
     
      \begin{enumerate}
      \item $T(L_i) = W_i$,
     \item if $s\notin \bigcap_{i\in I} W_i$ and $s \to_{L_i} t$ 
     then $t \not\to_{L_j} s$ for all $j\neq i$.
 \end{enumerate}
\end{definition}

As to the guidance relations $\to_{L_Y}$ and $\to_{L_Z}$, 
 for every $s\in S$, $s\to_{L_Y} s'$ and $s\to_{L_Z} s''$, where  
\[s': = \begin{cases}
    s & \text{if } \overline{y}_s = \overline{0}\\
    \overline{0}\overline{z'} & \text{if } \overline{y}_s \neq \overline{0}
\end{cases}\quad\quad  s'' := \begin{cases}
    s & \text{if } \overline{z}_s = \overline{0}\\
   \overline{y''} \overline{0} & \text{if } \overline{z}_s \neq \overline{0}
\end{cases} \]
and, for every $1\leq i\leq 3$, $z'(i) = 1$ iff $y_s(j) = 1$ and  $y_j\to_Y z_i$ for some $1\leq  j\leq 5$, and for every $1\leq i\leq 5$, $y''(i) = 1$ iff $z_s(j) = 1$ and  $z_j\to_Z y_i$ for some $1\leq  j\leq 3$. 

Hence, $\to_{L_Y}$ guides any profile $s\in S^P$ to itself if the value of all its $y$-coordinates is $0$ (i.e.~if $s\in W^P_Y$), and otherwise, to an element $s'\in W^P_Y$  whose $z$-coordinates take the values determined by which   $y$-coordinate  of $s$ is non-zero, and to which $z$-variable the $y$-variable corresponding to that coordinate is mapped  by the original relation $\to_Y$, and symmetrically for $\to_{L_Z}$. For instance, if $s = \overline{y}_s\overline{z}_s$ with $\overline{y}_s = (1, 1, 0, 0, 1)$ and $\overline{z}_s = \overline{0}$, then $s' = (0, 0, 0, 0, 0, 1, 1, 0)$, since $y_1\to_Y z_1$, $y_2\to_Y z_2$, and $y_5\to_Y z_1$.

\begin{proposition}
    The codices $(L_Y, L_Z)$ magically implement $(S^P, W^P_Y, W^P_Z)$.
\end{proposition}
\begin{proof}
    By construction, $W^P_U : = T(L_U)$ for each $U\in \{Y, Z\}$. To finish the proof, we need to show that, for any $s\in S^P$,  $s\neq \overline{0}$, if  $s\to_{L_Y} s'$, then $s'\not\to_{L_Z} s$ and if $s\to_{L_Z} s''$, then $s''\not\to_{L_Y} s$. By assumption, $s$ has at least one non-zero coordinate. We proceed by cases. If $\overline{y}_s = \overline{0}$, then $s' = s$, and the non-zero coordinates in $s' = s$ are among the $z$-coordinates; however, by definition, the element $s''\in S$ s.t.~$s'\to_{L_Z} s''$ is such that $\overline{z}_{s''} = \overline{0}$, hence  $s''\neq s$, and so $s'\not\to_{L_Z} s$. If $\overline{y}_s \neq \overline{0}$, then $s': = \overline{0}\overline{z'}$, where $\overline{z'}\neq \overline{0}$. To finish the proof, we need to show that  if $s''\in S$ and $s'\to_{L_Y} s''$, then $s''\neq s$. Since $\overline{z}_{s'} = \overline{z'}\neq \overline{0}$, by definition, $s'' = \overline{y''}\overline{0}$. Hence, if $\overline{z}_s\neq \overline{0}$, we can conclude that $s''\neq s$. Let us assume that  $\overline{z}_s = \overline{0}$. Then, the only non-zero coordinates in $\overline{z}_{s'}$ are those that correspond to the variables $z_i$ such that $y_j\to_{Y}z_i$ for some non-zero $y$-coordinate $y_j$ in $s$. As discussed at the end of Section \ref{ssec: magical}, the relations $\to_Y$ and $\to_Z$ magically implement the original implementation problem $(S, W_Y, W_Z)$; hence, $y_j\to_{Y}z_i$ implies  $z_i\not\to_{Z}y_j$ for any non-zero coordinate of $\overline{z}_{s'}$. This implies that the $j$th coordinate of $\overline{y}_{s''} = \overline{y''}$ is zero, and hence $s''\neq s$, as required. 
\end{proof}
Notice that, although the construction presented above has been introduced in the context of the example discussed at the end of Section \ref{ssec: magical}, the definitions and the proof above do not depend on the specific example, and hence have general validity.

\section{ Tiglath-Pileser relations and game-theoretic analyses}
\label{sec:TP relations and game-theoretic}

In the present section, we revisit the  formal  frameworks of \cite{glazer2012model, glazer2024magical}, reported on in Section \ref{sec: preliminaries}, by framing them in a purely set-theoretic way. 
Specifically, in Section  \ref{ssec:models of persuasion via TP relations}, we   isolate a class of  binary relations (which we refer to as  {\em Tiglath-Pileser} relations)  that  can function as guidance relations, and use them  in the context of a game-theoretic  analysis of the persuasion situations of \cite{glazer2012model};
in Section \ref{ssec:Magical implementation via relations},  we pursue an axiomatic analysis on Tiglath-Pileser relations  that  can  function as `magical implementers'.  A common feature of these developments is that the analysis is structured along a hierarchy of levels of strategic/cognitive awareness of the agents involved.  

\subsection{Models of persuasion via Tiglath-Pileser relations} \label{ssec:models of persuasion via TP relations}

In this section, as in \cite{glazer2012model},  we consider persuasion  situations  with a Speaker $A$ and Listener $B$, in which $A$ makes declarations  aimed at persuading $B$ of a certain state of affairs, irrespective of the true state of affairs, whereas $B$  tries to guess the true  state of affairs using $A$'s message. We  represent these persuasion situations as  tuples $(S, W_A, R_A)$ such that $S$ is the set of states,   $W_A \subseteq S$ 
is the winning set of $A$ and $R_A \subseteq S \times S$ is a Tiglath-Pileser relation over $S$ towards $W_A$:

\begin{definition}\label{def:TP:rel}
For any tuple $(S, W)$ such that $S$  is a nonempty set and $W\subseteq S$,
    a {\em Tiglath-Pileser relation over $S$ towards $W$} is a  relation $R\subseteq S\times S$  verifying the following conditions:
    \begin{enumerate}
     \item[$\mathrm{R1}$.] $R^{-1} [S]: = \{s\in S\mid \exists s'(s'\in S\ \& \ R(s, s'))\} = S$; 
        \item[$\mathrm{R2}.$] $R_{|W\times S} = \Delta_{W}$; 
        \item[$\mathrm{R3}.$] $\forall s(s\notin W\Rightarrow R[s]\subseteq W \text{ or } R[s] = \{s\})$. 
    \end{enumerate}
    As in Definition \ref{def:pre-guiding relation}, for each such relation, we let $T(R) = \{s\in S\mid \forall t(R(s, t)\Rightarrow s = t)\}$.
    We let $\mathsf{TP}(S, W)$ denote the set of Tiglath-Pileser relations over $S$ towards $W$.
\end{definition} 
Tiglath-Pileser relations are understood as generalizations of   the guidance relations of Definition \ref{def:guidance}, and fulfill a similar function in modelling persuasion scenarios.
Condition R1 states that every $s\in S$ is $R$-related  to one or more states; hence, in every possible true state, the Tiglath-Pileser relation guides the speaker to some state; by R2, when the true state is in their own winning set, according to their Tiglath-Pileser relation, the agent is to declare it truthfully; by  R3, when the true state is not in their winning set, the Tiglath-Pileser relation will either let the agent declare some state in their winning set, or the true state.



 The game proceeds as follows: the Speaker announces a nonempty  $S' \subseteq S$, and the Listener concludes a subset $X \subseteq S$. 
The payoff of the Speaker is $1$ if $X \subseteq W_A$, and $0$ otherwise.\footnote{Graded payoffs $|W_A \cap X| \div |X|$ can  alternatively be attributed to the Speaker by considering the non-empty intersection of $X$ with $W_A$ as a partial win.}
The payoff of the Listener is $|\{s\} \cap X| \div |X|$, where  $s \in S$ denotes the true state. 

We  analyse situations in which the Speaker and the Listener display various degrees of strategic/cognitive awareness (see  \cite{camerer2004cognitive});  
specifically, we assume that  a non-strategic   $A$  always follows their  Tiglath-Pileser relation and announces some nonempty $S'\subseteq R_A[s]$, 
whereas a non-strategic   $B$ believes  $A$ to be honest.  We also assume that this is known by a strategic $B$ who furthermore knows the $R_A$ and $W_A$ associated with  a non-strategic $A$, and that all this is known by a strategic $A$ who knows that $B$ is strategic. 

We  consider the following   cases, corresponding to five levels of analysis, which we express in the language of epistemic logic, in which, for any $i\in \{A, B\}$, the proposition  $p_i$  stands for  `agent $i$ is strategic', and the epistemic operator $\Box_i$   stands for  `agent $i$ believes that'.

\begin{center}
    \begin{tabular}{c|l}
    Case 0    & $\neg p_A, \neg p_B$  \\
  Case 1     & $p_A, \neg p_B, \Box_A\neg p_B$  \\
 Case 2     & $\neg p_A,  p_B, \Box_B\neg p_A  $  \\
  Case 3     & $ p_A,  p_B, \Box_B\neg p_A, \Box_A p_B,\Box_A\Box_B\neg p_A$  \\
   Case  4     & $ p_A,  p_B, \Box_B p_A, \Box_A p_B,\Box_B\Box_A p_B, \Box_A\Box_B\neg p_A, \Box_B\Box_A\Box_B\neg p_A$   \\
    \end{tabular}
\end{center}


\paragraph{Case 0.}If $R_A[s] \cap W_A \neq \varnothing$, then $A$ announces $S' := R_A[s] \cap W_A$. Since $B$ is non-strategic, they take $A$ to be honest and thus believe the true state lies in $S'$. As $S' \subseteq W_A$, $A$ successfully persuades $B$. If $s \notin W_A$, then $s \notin S'$, and hence $B$'s payoff is $0$ and $A$'s payoff is $1$. 
If $s \in W_A$, then, by Definition~\ref{def:TP:rel}.2, $R_A[s] = S' = \{s\}$, which $B$ believes; thus the payoff of both players is $1$.
If $R_A[s] \cap W_A = \varnothing$, then by Definition~\ref{def:TP:rel}.3, $R_A[s] = \{s\}$. Hence, $A$ is forced to announce $S' = \{s\}$, which results in $A$'s payoff being $0$ while $B$'s payoff is $1$.   


\paragraph{Case 1.}
Since $A$ is strategic and knows that $B$ is non-strategic, $A$ can announce any nonempty $S' \subseteq  W_A$ (independently of $R_A$), and as $B$ is non-strategic, they  believe $A$ who then gets a payoff of $1$. If $s \notin S'$, then $B$'s payoff is $0$; if $s \in S'$, then $B$'s payoff is  $1/|S'|>0$.


\paragraph{Case 2.}
In this case, $A$ is non-strategic, while   $B$ is strategic, knows that $A$ is non-strategic, and knows $R_A$.
Then $A$ announces $S' := R_A[s] \cap W_A$ if $R_A[s] \cap W_A \neq \varnothing$, and $S' := R_A[s]= \{s\}$ otherwise, and $B$ concludes $X: =  R_A^{-1}[S']$; by construction,  $s\in   X$, which guarantees a positive payoff for $B$, while $A$'s graded payoff is nonzero only if $X\cap W_A = R^{-1}_A[S']\cap W_A  \neq\varnothing$ iff $S'\cap R_A[W_A] = S'\cap W_A\neq \varnothing$
iff $R_A[s]\cap W_A \neq \varnothing$. If $R_A[s]\cap W_A  = \varnothing$ then $B$'s payoff is $1$ and $A$'s payoff is $0$.

\paragraph{Case 3
.}
In this case, $A$ knows that for any nonempty $S' \subseteq S$ they announce, $B$ will conclude  $X: = R_A^{-1}[S']$ as in Case 2. Hence, $A$ declares a nonempty $S' \subseteq S$ that maximises 
$A$'s graded payoff, which is   $|R_A^{-1}[S'] \cap W_A|/ |R_A^{-1}[S']| $. 
If some $S'$ exists s.t.~$R_A^{-1}[S'] \subseteq W_A$, then $A$ can achieve a payoff of $1$  by declaring such an $S'$. Otherwise, since $W_A \subseteq R_A^{-1}[W_A]$, $A$ can secure a graded  payoff of $|W_A|/|R_A^{-1}[W_A]|$ by declaring $S':=W_A$.

\paragraph{Case 4
.}
As in the previous case,  if  some   $S'$ exists s.t.~$R_A^{-1}[S'] \cap W_A \neq \varnothing$ $A$ announces one    that maximizes $|R_A^{-1}[S'] \cap W_A|/ |R_A^{-1}[S']|$; otherwise, $A$ announces  $S' := \{s\}$.
Here, $B$ can infer the true state only if no such $S'$ exists; then $S'=\{s\}$ and $B$ trivially identifies $s$. If such an $S'$ exists, $A$ may announce it regardless of the true state, and $B$ knows this; hence $B$ cannot infer anything from the message. Thus,  $A$ cannot persuade  $B$, while $B$ gains no information about the true state.

\subsection{Magical implementation via relations}\label{ssec:Magical implementation via relations} 
As in \cite{glazer2012model, glazer2024magical}, we consider situations in which one or more Speakers make declarations  aimed at persuading a Listener of a certain state of affairs, irrespective of the true state of affairs. We assume that the Speakers' declarations are not arbitrary, but  are always anchored to the true state of affairs, because the Speakers are either limited in their ability to lie by the Listener's knowledge and/or their own bounded rationality. We represent these {\em persuasion situations} as tuples     
 $(S, \{W_i\}_{i\in I}, \{R_i\}_{i\in I})$ s.t.~$S$  is a nonempty domain of states, $I$ is a nonempty finite set of agents,  $W_i\subseteq S$, and each $R_i$ is a  Tiglath-Pileser relation on $S$ towards $W_i$ for all $i\in I$ (cf.~Definition \ref{def:TP:rel}). We do not insist on requiring that $\{W_i\}_{i\in I}$ be a partition of $S$; hence, we allow for more than one Speaker to be truthful, or none.
Intuitively, agent $i$ wishes to persuade the Listener that the true state belongs to  $W_i$ (the {\em winning set} of $i$). 
Hence, if $t\in S$ is the true state, agent $i$ will declare some $s_i\in R_i[t]: = \{s\in S\mid R_i(t, s) \}$. Therefore, upon knowing $R_i$, the Listener can apply the Tiglath-Pileser principle to conclude that the true state is in $R_i^{-1}[s_i]: = \{s\in S\mid R_i(s, s_i) \}$ when $i$ declares $s_i$. Likewise, upon receiving  the declarations $s_i$ of each Speaker $i\in I$, the Listener will  conclude that  $t\in \bigcap_{i\in I}R_i^{-1}[s_i]$. 
In most situations, we will assume that Tiglath-Pileser relations satisfy a {\em stronger version} of R3 in Definition \ref{def:TP:rel}, namely $R_i[S] \subseteq W_i$. In this case, $T(R_i) = W_i$.  The next definition is the set-theoretic reformulation of Definition \ref{def: magical implementation extended}. 
\begin{definition}
    \label{def: magical implementation extended set-theoretic}
     The set of relations $\{R_i\mid i\in I\}$ {\em magically implements} $(S, \{W_i\mid i\in I\})$ if:   
      \begin{enumerate}
      \item $R_i[S] \subseteq W_i$ and $R_{i|W_i\times W_i} = \Delta_{W_i}$ for all $i\in I$;
     \item for all $s, t \in S$, if $s\notin \bigcap_{i\in I} W_i$ and $t\in R_i[s]$ , 
     then $s \notin R_j[t]$ for all $j\neq i$.
 \end{enumerate}
\end{definition}
 For simplicity of presentation, we consider only two agents, $A$ and $B$, although the same analysis and constructions can be carried out for any finite number of agents.



For a persuasion situation $(S, W_A, W_B, R_A, R_B)$ as above, in which $R_A$ and $R_B$ satisfy R1, R2 and the stronger version of R3,
let $R\subseteq S\times (W_A\times W_B)$ be the {\em product relation} defined as follows: for all $s\in S$, $s'\in W_A$, and $s''\in W_B$,
\begin{equation}
\label{eq:product relation}
    R(s, (s', s'')) \quad \text{ iff }\quad R_A(s, s')\ \text{ and }\ R_B(s, s'').
\end{equation}
Hence, by definition, for all $s\in S$, $s'\in W_A$ and $s''\in W_B$,  \[R[s] = R_A[s]\times R_B[s]\quad \text{  and }\quad  R^{-1}[(s', s'')] = R_A^{-1}[s']\cap R_B^{-1}[s''].\]
  Assumption R1 ($R^{-1}_A [W_A] = S = R^{-1}_B [W_B]$) guarantees that, whenever $A$ and $B$ are not {\em strategic} (in the sense that neither  takes the other parties' actions and knowledge into consideration), for every ``declaration'' $(s', s'')\in W_A\times W_B$ from  $A$ and $B$, the principal can always get back to the set of states $s\in R^{-1}[(s', s'')]$  from which the declarations possibly originated (this formally represents the `anchor to the truth'). This is because, when both $A$ and $B$ are not strategic and $t\in S$ is the true state, $t\in R_A^{-1}[s']\cap R_B^{-1}[s''] = R^{-1}[(s', s'')]$. Furthermore, if 
  
\begin{enumerate}
    \item[R4.] $R$ is {\em relation-injective}, i.e.~$s\neq t$ implies   $R[s]\cap R[t] = \varnothing$ for all $s, t\in S$,\footnote{$R\subseteq X\times Y$ is relation-injective iff $R^{-1}\subseteq Y\times X$ is functional (that is, for all $y\in Y$, $R^{-1}[y]$  is either the empty set or is a singleton).  Indeed, if $R^{-1}$ is functional and $y\in Y$ s.t.~$y\in R[x']\cap R[x'']$ i.e.~$x', x''\in R^{-1}[y]$, then, by functionality, $x' = x''$. Conversely, if $x', x''\in R^{-1}[y]$, i.e.~$y\in R[x']\cap R[x'']$, by relation-injectiveness $x' = x''$, as required. A sufficient condition for the product relation $R$ to be relation-injective is if either $R_A$ or $R_B$ is; however, as we will see in Example \ref{example:magical:relations}, this condition is not necessary.}
\end{enumerate}
    then for all $s'\in W_A$ and $s''\in W_B$, the set $R^{-1}[(s', s'')]$ has at most one element (for if $R(s, (s', s''))$ and $R( t, (s', s''))$, then $(s', s'')\in R[s]\cap R[t]\neq \varnothing$), and hence, the Listener can always get back to the true state.

As we do not assume that $W_A\cap W_B = \varnothing$, if $t$ belongs to both winning sets, then both agents would truthfully declare it, and in fact, they would agree only on states in the intersection of their winning sets.

\begin{proposition}
\label{prop:magical implementation and relation injectiveness}
 The following are equivalent for any  $(S, W_A, W_B, R_A, R_B)$ satisfying the condition of Definition \ref{def: magical implementation extended set-theoretic}.1 and such that $W_A\cup W_B=S$:
    \begin{enumerate}
        \item 
     $\{R_A, R_B\}$ magically implements $(S, W_A, W_B)$; \item the product relation $R$ defined as in \eqref{eq:product relation} is relation-injective.
     \end{enumerate}
\end{proposition}
\begin{proof}
It is enough to show that   Definition \ref{def: magical implementation extended set-theoretic}.2 is equivalent to $R$ being relation-injective while assuming  the condition of Definition \ref{def: magical implementation extended set-theoretic}.1. 
Arguing contrapositively, let us assume that $R$ is not relation-injective, i.e.~some $s_1, s_2, s, t\in S$ exist such that
 $s_1\neq s_2$, $R_A(s_i,s)$ and $R_B(s_i,t)$ for $1\leq i\leq 2$. Hence, $s\in W_A$ and $t\in W_B$ by  Definition \ref{def: magical implementation extended set-theoretic}.1. Since $W_A\cup W_B=S$,  we can assume without loss of generality that $s_1\in W_A$  and $s_2\notin W_A$ (hence, $s_2\in W_B$); this is because, if e.g.~$s_1,s_2\in W_A$, then $s_1 = s = s_2$ by  Definition \ref{def: magical implementation extended set-theoretic}.1, against the assumption that $s_1\neq s_2$, and a similar argument excludes $s_1, s_2\in W_B$.  Therefore, $R_A(s_1,s)$ and $R_B(s_2, t)$ respectively imply $s_1=s$ and $s_2 = t$ by  Definition \ref{def: magical implementation extended set-theoretic}.1.   Then, $s_2 \notin W_A\cap W_B$,  $s_1\in R_A[s_2]$ and $s_2\in R_B[s_1]$, which violates Definition \ref{def: magical implementation extended set-theoretic}.2. 

Conversely, if $s\notin W_B$ and $s\in R_A[t]$ and $t\in R_B[s]$, then $t\in W_B$, so $t\neq s$. Moreover, the assumptions and   Definition \ref{def: magical implementation extended set-theoretic}.1 imply that $R(s,(s,t))$ and $R(t,(s,t))$, i.e.\ $R$ is not relation-injective.
\end{proof}

\begin{example}
    The last example discussed at the end of Section \ref{ssec: magical} gives rise to the persuasion situation $(S, W_Y, W_Z, \to_Y, \to_Z)$ represented below, which satisfies all the assumptions of Proposition \ref{prop:magical implementation and relation injectiveness}. Hence, an equivalent way of understanding why   $\to_Y$ and $\to_Z$  magically implement $(S, W_Y, W_Z)$ is that, whichever state is the true one, the principal can consider the inverse image of the tuple declared by $Y$ and $Z$ according to their  product relation $R\subseteq S\times (W_Y\times W_Z)$, defined as in \eqref{eq:product relation} and represented in the third column below. Since $R$ is relation-injective, the inverse image of each tuple $(s', s'')\in W_Y\times W_Z$ has at most one element; when such an element exists, under the assumption that neither Speaker is strategic, the principal can conclude that it is the true state.

    \begin{center}
    \begin{tabular}{c| c c | c}
    \Huge{$\to_{Y}$} & \Huge{$\to_{Z}$} && \Huge{$R$}\\
    & &\\

\begin{tikzpicture}[baseline=(current bounding box.north), scale=1]  

\node (z1) at (0,3) {};
\node (z2) at (0,2) {};
\node (z3) at (0,1) {};
\node (y1) at (2,3) {};
\node (y2) at (2,2) {};
\node (y3) at (2,1) {};
\node (y4) at (2,0) {};
\node (y5) at (2,-1) {};

\draw[<-] (z1) -- (y1);
\draw[<-] (z2) -- (y2);
\draw[<-] (z3) -- (y3);
\draw[<-] (z1) -- (y4);
\draw[<-] (z2) -- (y5);

\node[anchor=east] at (z1) {$z_1$};
\node[anchor=east] at (z2) {$z_2$};
\node[anchor=east] at (z3) {$z_3$};
\node[anchor=west] at (y1) {$y_1$};
\node[anchor=west] at (y2) {$y_2$};
\node[anchor=west] at (y3) {$y_3$};
\node[anchor=west] at (y4) {$y_4$};
\node[anchor=west] at (y5) {$y_5$};


 \filldraw[thick, purple, opacity=0.2] (-0.5,0) rectangle (0.5, 3.5);
\end{tikzpicture}      & 
\begin{tikzpicture}[baseline=(current bounding box.north), scale=1]  

\node (z1r) at (4,3) {};
\node (z2r) at (4,2) {};
\node (z3r) at (4,1) {};
\node (y1r) at (6,3) {};
\node (y2r) at (6,2) {};
\node (y3r) at (6,1) {};
\node (y4r) at (6,0) {};
\node (y5r) at (6,-1) {};

\draw (z1r)[->] -- (y2r);
\draw (z2r)[->] -- (y3r);
\draw (z3r)[->] -- (y4r);

\node[anchor=east] at (z1r) {$z_1$};
\node[anchor=east] at (z2r) {$z_2$};
\node[anchor=east] at (z3r) {$z_3$};
\node[anchor=west] at (y1r) {$y_1$};
\node[anchor=west] at (y2r) {$y_2$};
\node[anchor=west] at (y3r) {$y_3$};
\node[anchor=west] at (y4r) {$y_4$};
\node[anchor=west] at (y5r) {$y_5$};

\filldraw[thick, blue, opacity=0.2] (5.5,-1.3) rectangle (6.5, 3.5);
 
\end{tikzpicture}

&& 
\begin{tabular}[t]{c|c}
true state & declarations\\
$y_1$ & $(z_1, y_1)$  \\
$y_2$ & $(z_2, y_2)$  \\
$y_3$ & $(z_3, y_3)$  \\
$y_4$ & $(z_1, y_4)$  \\
$y_5$ & $(z_2, y_5)$  \\
$z_1$ & $(z_1, y_2)$  \\
$z_2$ & $(z_2, y_3)$  \\
$z_3$ & $(z_3, y_4)$  \\
\end{tabular}
\\
     
    \end{tabular}
\end{center}
\end{example}

\begin{example}\label{example:magical:relations}
The present example  shows that the magical implementation mechanism can work also under more general  assumptions than those of \cite{glazer2024magical}; for instance, the winning sets do not need to form a partition. Compared to the previous example, the present one also shows that the individual Tiglath-Pileser relations do not need to be  relation-injective in order for their product to be. In this more general case, the pivot is not anymore Definition \ref{def: magical implementation extended set-theoretic}, but rather  conditions $\mathrm{R1}$-$\mathrm{R4}$.
Let $S = \{a_1, a_2, a_3, b_1, b_2, b_3, s \}$, $W_A = \{a_1, a_2, a_3\}$, $W_B = \{b_1, b_2, b_3\}$ (hence, $W_A\cup W_B\neq S$),   and let $R_A$ and $R_B$ be as in the pictures below (except that ${R_A}_{|W_A\times S} = \Delta_{W_A}$ and ${R_B}_{|W_B\times S}= \Delta_{W_B}$ are not represented in the pictures). The ensuing relation $R\subseteq S\times (W_A\times W_B)$ is represented in the third column below.

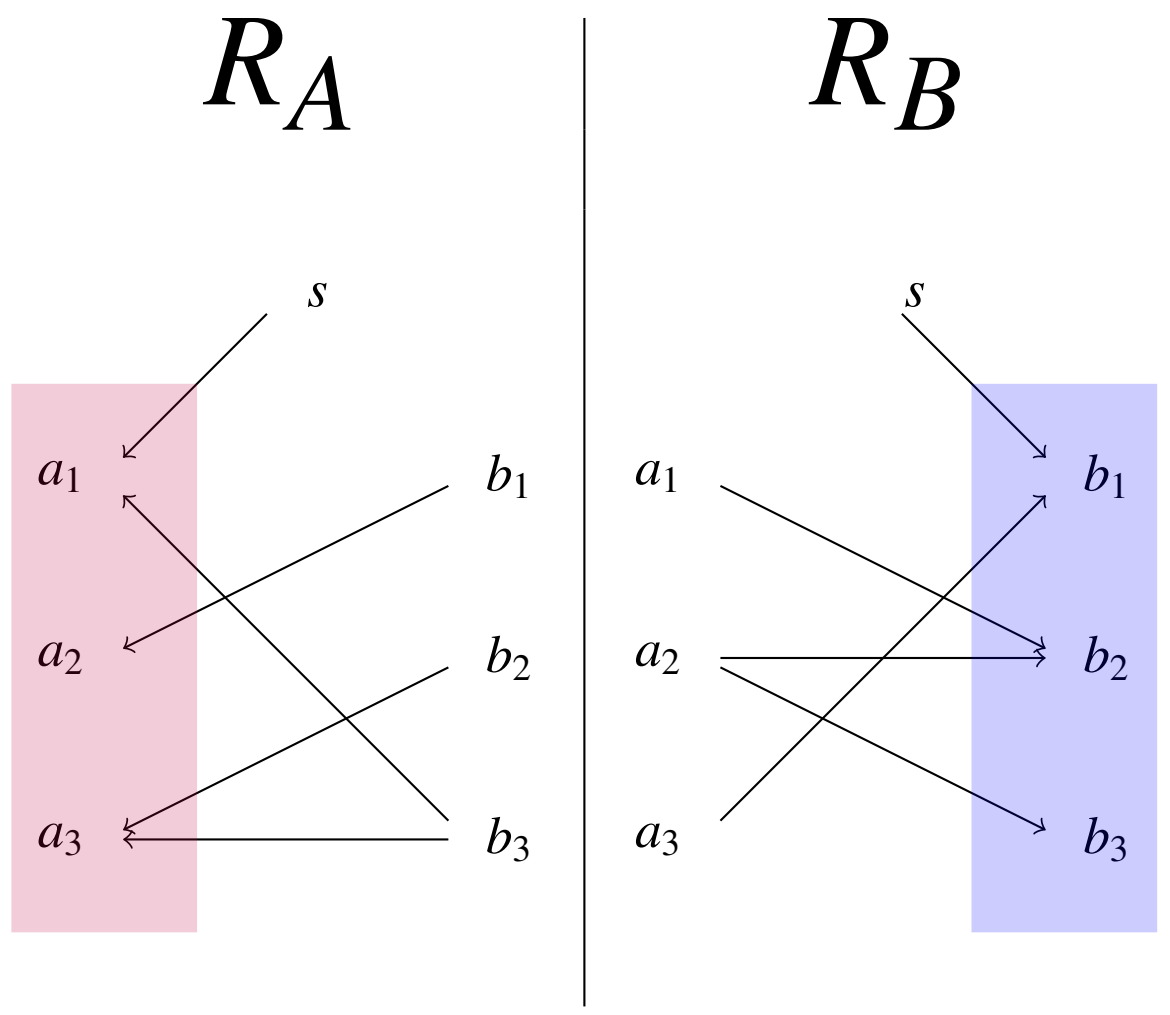
\begin{center}
\begin{tabular}{c|c c |c}
\Huge{$R_{A}$} & \Huge{$R_{B}$} & & \Huge{$R$} \\
& &&\\

\begin{tikzpicture}[baseline=(current bounding box.north), scale=1.2]

\node (a1) at (0,3) {};
\node (a2) at (0,2) {};
\node (a3) at (0,1) {};
\node (b1) at (2,3) {};
\node (b2) at (2,2) {};
\node (b3) at (2,1) {};
\node (s) at  (1,4) {};

\draw[->] (b1) -- (a2);
\draw[->] (b2) -- (a3);
\draw[->] (b3) -- (a1);
\draw[->] (b3) -- (a3);
\draw[->] (s) -- (a1);

\node[anchor=east] at (a1) {$a_1$};
\node[anchor=east] at (a2) {$a_2$};
\node[anchor=east] at (a3) {$a_3$};
\node[anchor=west] at (b1) {$b_1$};
\node[anchor=west] at (b2) {$b_2$};
\node[anchor=west] at (b3) {$b_3$};
\node[anchor=west] at (s) {$s$};

\filldraw[thick, purple, opacity=0.2] (-0.5,0.5) rectangle (0.5, 3.5);
\end{tikzpicture}
&
\begin{tikzpicture}[baseline=(current bounding box.north), scale=1.2]

\node (a1r) at (4,3) {};
\node (a2r) at (4,2) {};
\node (a3r) at (4,1) {};
\node (b1r) at (6,3) {};
\node (b2r) at (6,2) {};
\node (b3r) at (6,1) {};
\node (sr) at (5,4) {};

\draw (a1r)[->] -- (b2r);
\draw (a2r)[->] -- (b2r);
\draw (a2r)[->] -- (b3r);
\draw (a3r)[->] -- (b1r);
\draw (sr) [->] -- (b1r);

\node[anchor=east] at (a1r) {$a_1$};
\node[anchor=east] at (a2r) {$a_2$};
\node[anchor=east] at (a3r) {$a_3$};
\node[anchor=west] at (b1r) {$b_1$};
\node[anchor=west] at (b2r) {$b_2$};
\node[anchor=west] at (b3r) {$b_3$};
\node[anchor=west] at (sr) {$s$};

\filldraw[thick, blue, opacity=0.2] (5.5,0.5) rectangle (6.5, 3.5);
\end{tikzpicture}
&&
\begin{tabular}[t]{c|c}
true state & declarations\\
$s$ & $(a_1, b_1)$  \\
$a_1$ & $(a_1, b_2)$  \\
$b_1$ & $(a_2, b_1)$  \\
$a_2$ & $(a_2, b_2)$  \\
$a_2$ & $(a_2, b_3)$  \\
$b_2$ & $(a_3, b_2)$  \\
$a_3$ & $(a_3, b_1)$  \\
$b_3$ & $(a_1, b_3)$  \\
$b_3$ & $(a_3, b_3)$  \\
\end{tabular}
\\
\end{tabular}
\end{center}
The relations $R_A$ and $R_B$ satisfy $\mathrm{R1}$, $\mathrm{R2}$, and the stronger version of $\mathrm{R3}$, and $R$ satisfies  $\mathrm{R4}$. Moreover, $R$ is  onto $W_A \times W_B$ and  is not functional, since $(b_3, (a_1, b_3)), (b_3, (a_3, b_3))\in R$. Removing  $(a_2, b_3)$ from $R_B$ would result in removing $(a_2, b_3)$ from the image of   $R$, which would not be onto $W_A\times W_B$ any longer, while $R_B$ and $R$ would retain all the remaining properties.
   
\end{example}

Notice that, for R4 to be satisfied,  the pigeonhole principle requires that $|S| \leq |W_A \times W_B|$, which may be impossible if the size of $W_A$ and/or $W_B$ is small relative to $S$. For example, if $S = \{s_1, s_2, s_3, s_4\}$ and $W_A = \{s_1\}$ and $W_B = \{s_2, s_3, s_4\}$, then  $|S| = 4$ but $|W_A \times W_B| = 3$.

\paragraph{Strategic/cognitive hierarchy.} 
As done in Section \ref{ssec:models of persuasion via TP relations}, we  analyse situations in which agents display various degrees of strategic/cognitive awareness (see  \cite{camerer2004cognitive})  which allows them to depart from the basic assumption that their declarations will adhere to what their Tiglath-Pileser relation prescribes; in other words, their strategies may take into account the expectations of the other agents and of the principal revolving around their Tiglath-Pileser relations.
We discuss the following cases; in each of them, the principal's approximation of the true state obtains as the $R$-inverse image of the  tuple of states declared by $A$ and $B$.\footnote{The principal can then draw further conclusions in line with their position in the strategic/cognitive hierarchy. We will not pursue a detailed analysis of the principal's reasoning but rather focus on the agents' behaviour.} 
\begin{enumerate}
    \item Neither $A$ nor $B$ are strategic; they declare what their associated Tiglath-Pileser relation indicates;
    \item Agent $A$ is not strategic and declares what their associated Tiglath-Pileser relation indicates, while $B$ is strategic, assumes $A$ is not, and knows $A$'s Tiglath-Pileser relation;
    \item Both $A$ and $B$ are  strategic, assume the other agent is not, and know the other's Tiglath-Pileser relation;
    \item Both agents are strategic and know the other's Tiglath-Pileser relation, but while $A$ knows that $B$ is strategic, $B$ assumes that $A$ is not strategic;
    \item Both $A$ and $B$ are  strategic, assume the other agent is also strategic, and know the other's Tiglath-Pileser relation.
\end{enumerate}
In the \underline{first case}, the principal always succeeds in  restricting the search for the true state to  $R^{-1}[(s', s'')]$, and if $R$ is relation-injective (R4), the principal is able to establish the true state, and the agent $Y\in\{A, B\}$ s.t.~$s\in W_Y$  will also win.

In the \underline{second case}, if 
\begin{enumerate}
    \item[R5.] ${R_A}_{|(S\setminus (W_A \cup W_B))\times W_A}$ is functional, and symmetrically ${R_B}_{|(S\setminus (W_A \cup W_B))\times W_B}$ is functional (notice that this condition is trivially  satisfied when $W_A\cup W_B = S$)
    \item[R6.]  $R_A^{-1}[s]\cap W_B\neq \varnothing$ for all $s\in W_A$, and    $R_B^{-1}[s]\cap W_A\neq \varnothing$ for all $s\in W_B$, i.e.~${R_A}_{|{W_B\times W_A}}$ is onto $ W_A$, and ${R_B}_{|{W_A\times W_B}}$ is onto $ W_B$.
\end{enumerate}
then $B$ always wins: if $t\in W_B$, then  the pair of agents declares $(s', t)$ with $s'\in R_A[t]$, from which the principal correctly concludes that the true state $t$ is in $R^{-1}[(s', t)]$; if $R$ is relation-injective, the principal will identify $t$ as the only element in $R^{-1}[(s', t)]$. If $t\notin W_B$, then, by R5,   $B$ knows that $A$ will declare the unique $s'\in R_A[t]$ (notice that, by R2 and R5, $s'$ is unique both if $t\in W_A$ and if $t\notin W_A$); then $B$ declares some $s''\in W_B\cap R^{-1}_A[s']$ which exists since, by R6, $W_B\cap R^{-1}_A[s']\neq\varnothing$. Then the principal will consider the set $R^{-1}[(s', s'')] = R^{-1}_A[s']\cap R_B^{-1}[s'']$. By construction, $s''\in R^{-1}_A[s']\cap R_B^{-1}[s'']$, and since $R$ is relation-injective, $s''$ is the only element of $R^{-1}[(s', s'')]$. Hence, in this case, $B$ successfully deceives the principal into concluding that the true state is $s''$.

\begin{remark}
    Conditions R5 and R6 are satisfied by the relations $R_A$ and $R_B$ described in Example \ref{example:magical:relations}. However, removing the pair $(a_2, b_3)$ from $R_B$, as discussed there, causes $R_B$ no longer to satisfy R6. Hence, there might be a tension between the non-surjectivity of $R$ and R6.
\end{remark}
\begin{remark}
\label{rem: case R4 prime}
In case R5 does not hold,  knowing $R_A$ is in general not sufficient for agent $B$ to know what agent $A$ will declare in the case that $s \not\in W_A$, which therefore makes it impossible to reliably pick the correct value $W_B\cap R^{-1}_A[s]$ to ensure that the principal concludes in $B$'s favour. However, the following  alternative assumption  ensures that there is an element in $W_B$ that covers every possible declaration $A$ could make, thanks to which  we can still make sure that $B$ wins using it and R6.
\begin{enumerate}
    \item [$\mathrm{R}5'$.]  for all $s \in S \setminus (W_A \cup W_B)$ some 
    $s' \in W_A$ exists such that $R_B[s]\subseteq R_B[s']$ and, symmetrically, 
    some $s''\in W_B$ exists such that $R_A[s]\subseteq R_A[s'']$. 
\end{enumerate}
Indeed, if $s\in W_A$, then we proceed as indicated above, using R6 and the fact that $B$ knows that $A$ will announce $s$;
if $s \in S \setminus (W_A \cup W_B)$, then, by $\mathrm{R}5'$,  agent $B$ can announce any  $s''\in W_B$ such that $R_A[s]\subseteq R_A[s'']$, while agent $A$ announces some $s' \in R_A[s]$. The principal considers $R^{-1}[(s', s'')] = R^{-1}_A[s']\cap R_B^{-1}[s'']$. From $s''\in W_B$, it follows that $s''\in R_B^{-1}[s'']$. From $s'\in R_A[s]\subseteq R_A[s'']$ it follows that $s''\in R_A^{-1}[s']$. Hence, $s''\in R^{-1}[(s', s'')]$, and since $R$ is relationally injective, the principal wrongly concludes that   $s''$ is the ``true'' state, and agent $B$ wins.
\end{remark}

In the \underline{third case} (recall that we have assumed that the true state $s\notin W_A\cap W_B$), we need to distinguish several subcases: if $s\in W_A\cup W_B$, then  the agent $Y\in\{A, B\}$ s.t.~$s\notin W_Y$  will win, since the two agents will behave as in the two subcases of the second case (each of the two agents will play it strategically but assuming that the other would not). The principal is always deceived by relying on $R$.

If $s\in S\setminus (W_A\cup W_B)$, then assuming, according to $\mathrm{R}5'$, that some $s'\in W_A$ exists s.t.\ $R_B[s]\subseteq R_B[s']$ and  some $s''\in W_B$ exists s.t.\ $R_A[s]\subseteq R_A[s'']$,  
we consider the following  sub-subcases: 
\begin{enumerate}
    \item[(a)] $s'\in R_A[s]$ and  $s''\in R_B[s]$;
     \item[(b)] $s'\in R_A[s]$ and  $s''\notin R_B[s]$;
      \item[(c)] $s'\notin R_A[s]$ and  $s''\in R_B[s]$;
       \item[(d)] $s'\notin R_A[s]$ and  $s''\notin R_B[s]$.
\end{enumerate}

If (a), then, despite their being strategic, $A$ and $B$ declare what they would have declared if they were not. Hence, by considering $R^{-1}[(s', s'')]$, the principal is able to retrieve the true state and wins.
Similarly,  if (b) or (c), then we would fall to the  case in which one agent is strategic and the other is not, that is, the strategic agent wins and the principal is tricked.
If (d), then more subcases:

\begin{enumerate}
    \item[(d1)] $s'\in R_A[s'']\setminus R_A[s]$ and  $s''\in R_B[s'] \setminus R_B[s]$;
     \item[(d2)] $s'\in R_A[s'']\setminus R_A[s]$ and  $s''\in W_B\setminus R_B[s']$;
      \item[(d3)] $s'\in W_A\setminus R_A[s'']$ and  $s''\in R_B[s']\setminus R_B[s]$;
       \item[(d4)] $s'\in W_A\setminus R_A[s'']$ and    $s''\in W_B\setminus R_B[s']$.
\end{enumerate}
Case (d1) cannot occur if $s'\neq s''$. Indeed,  the assumptions $s'\in R_A[s'']$ and $s''\in R_B[s']$ imply   $s''\in R_A^{-1}[s']$ and $s'\in R_B^{-1}[s'']$, and since assumption R2, i.e.\ that  ${R_A}_{|W_A\times W_A} = \Delta_{W_A}$ and  ${R_B}_{|W_B\times W_B} =  \Delta_{W_B}$, implies that $s'\in R^{-1}_A[s']$ and $s''\in R^{-1}_B[s'']$, it follows that $s', s''\in R^{-1}_A[s']\cap R^{-1}_B[s'']  = R^{-1}[(s', s'')]$, against the relation-injectivity of $R$.

If (d2), then the unique element in $R^{-1}[(s', s'')] = R^{-1}_A[s']\cap R^{-1}_B[s'']$   coincides with $s''$, since $s'\in R_A[s'']$ iff $s''\in R_A^{-1}[s']$. Hence, $B$ wins, and the principal and $A$ lose. The case of (d3) is symmetric, with $A$ winning, and $B$ and the principal losing.

If (d4), then, by relation-injectivity (R4),  $R^{-1}[(s', s'')] = R^{-1}_A[s']\cap R^{-1}_B[s'']$  contains at most one element. If this element exists, it cannot be $s''$, since, by assumption, $s'\notin R_A[s'']$ i.e.~$s''\notin R^{-1}_A[s']$, nor  can it be $s'$, since, by assumption, $s''\notin R_B[s']$ i.e.~$s'\notin R^{-1}_B[s'']$. So both $A$ and $B$ lose, but the principal does not know that they are strategic. If $R^{-1}[(s', s'')] = \varnothing$, then both $A$ and $B$ lose, and the principal knows that $A$ and $B$ deviated from their response relations.

\begin{figure}
    \centering
    \begin{center}
	\begin{tikzpicture}[scale=1]
		
		\draw (0,0) ellipse (5 and 4);
		\node at (5,2) {$S$};
		
		\draw (-2,0) ellipse (1.6 and 2);
		\node at (-3.3,1.8) {$W_A$};
		
		\draw (2,0) ellipse (1.6 and 2);
		\node at (3.3,1.8) {$W_B$};
		
		\draw (-2,0.7) ellipse (0.8 and 1.2);
		\node at (-3,-0.3) {{\footnotesize $R_A[s'']$}};
		
		\draw (-2,0.7) ellipse (0.4 and 0.6);
		\node at (-1.9,-0.1) {{\footnotesize $R_A[s]$}};
		
		\draw (2,0.7) ellipse (0.8 and 1.2);
		\node at (2.9,-0.3) {{\footnotesize $R_B[s']$}};
		
		\draw (2,0.7) ellipse (0.4 and 0.6);
		\node at (2.1,-0.1) {{\footnotesize $R_B[s]$}};
		
		\node[fill=black,circle,inner sep=1pt,label=above:$s$] at (0,0.7) {};
		\node[fill=black,circle,inner sep=1pt,label=below:$s'$] at (-2,-1.5) {};  
		\node[fill=black,circle,inner sep=1pt,label=below:$s''$] at (2,-1.5) {};  
		
		\draw (0,0.7) -- (-2,1.3);  
		\draw (0,0.7) -- (-2,0.1);  
		
		\draw (0,0.7) -- (2,1.3);  
		\draw (0,0.7) -- (2,0.1);  
		
		\draw[dashed] (-2,-1.5) -- (1.6,1.75);  
		\draw[dashed] (-2,-1.5) -- (2,-0.5);  
		
		\draw[dashed] (2,-1.5) -- (-1.6,1.75);  
		\draw[dashed] (2,-1.5) -- (-2,-0.5);  
		
	\end{tikzpicture}
	\end{center}
    \caption{An illustration of the case considered in Remark \ref{rem: case R4 prime}.}
    \label{fig:case:d4:relations}
\end{figure}
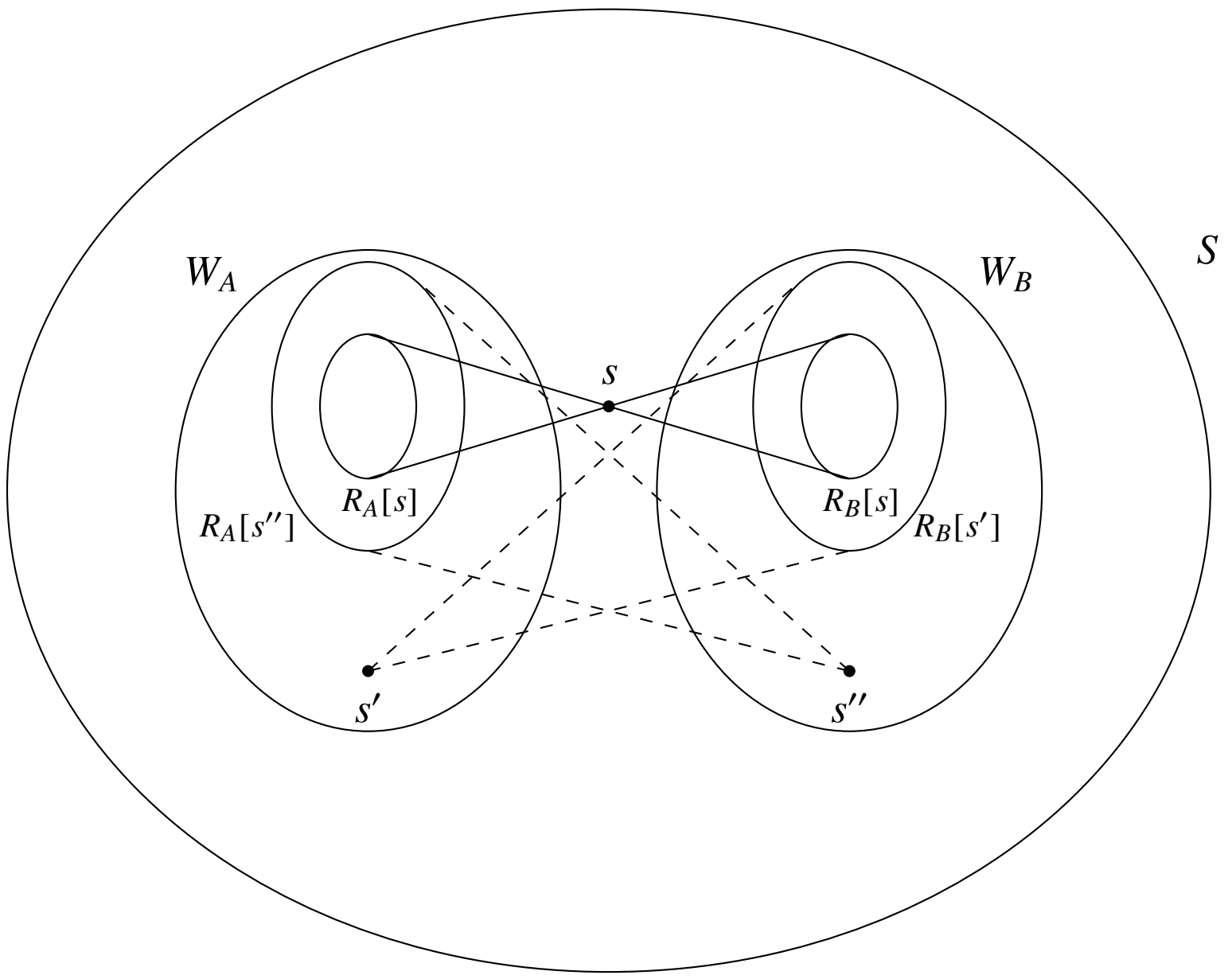

\paragraph{Taking stock.} For the principal, the best opportunity for detecting strategic play is if the range of $R$ is small relative to $W_A \times W_B$, since this would restrict the deviations that the agents could make from their response relations. Trying to minimize the range of $R$ is limited by the requirement that $R$ satisfy R4. Condition
R4 requires there to be at least as many pairs in $range(R)$ as there are elements in $S$. Hence, the number of pairs from $W_A \times W_B$ \emph{not} in the range of $R$ is at most $|W_A \times W_B| - |S|$. Therefore, for the principal, the optimal requirements for $R$ are R4 together with the following condition
\begin{enumerate}
    \item[R7.] the number of pairs from $W_A \times W_B$ \emph{not} in the range of $R$ is exactly  $|W_A \times W_B| - |S|$.
\end{enumerate}
 The best situation for the principal is therefore when $W_A$ and $W_B$ form a partition of $S$ into equally sized parts, if $S$ is even, or differing in size by at most one if $S$ is odd.   

In the \underline{fourth case}, we consider various subcases:  if $s\in W_B$, then $A$ wins by applying the same strategy as in the third case, namely, $B$, who believes that $A$ is non strategic, will announce  $s$ truthfully, hence $A$ will win by announcing an $s'\in R_B^{-1}[s]\cap W_A$, and the principal will erroneously conclude that the true state is $s'$. If $s\in W_A$, then $B$, who believes that $A$ will truthfully announce $s$ since $A$ (as $B$ believes) is non strategic, will announce some $s''\in R_A^{-1}[s]\cap W_B$; if 
\begin{enumerate}
    \item[R8.] $R_A^{-1}[s]\cap W_B = \{s''\} $ for all $s\in W_A$, and    $R_B^{-1}[s]\cap W_A = \{s'\}$ for all $s\in W_B$.
\end{enumerate}
then the state $s''$ announced by $B$ is unique, and hence $A$, by R6, can announce some $s'\in R_B^{-1}[s'']\cap W_A$ and wins, while the principal erroneously concludes that $s'$ is the true state.

If $s\in S\setminus (W_A\cap W_B)$, agent $B$, who believes that $A$ is not strategic, believes that $A$ will announce some $s'\in R_A[s]$, hence by $\mathrm{R}5'$, agent $B$ will announce some $s''\in W_B$ such that $R_A[s]\subseteq R_A[s'']$. If we further assume that the following strengthened version of $\mathrm{R}5'$ holds:
\begin{enumerate}
    \item [$\mathrm{R}5''$.]  for all $s \in S \setminus (W_A \cup W_B)$ some unique 
    $s' \in W_A$ exists such that $R_B[s]\subseteq R_B[s']$ and, symmetrically, 
    some unique $s''\in W_B$ exists such that $R_A[s]\subseteq R_A[s'']$.
\end{enumerate}
then $A$ announces some $s'''\in R_B^{-1}[s'']\cap W_A$ and win, while the principal erroneously concludes that $s'''$ is the true state.

\begin{remark}
\label{rem: case R4 prime}
In case $\mathrm{R}5'$ holds but its strengthening $\mathrm{R}5''$ does not,  $A$ can still win provided the following  alternative assumption  holds, which ensures that there is some element $s'''$ in $W_A$ that covers every possible declaration $B$ could make.
\begin{enumerate}
    \item [$\mathrm{R}5'''$.]  for all $s \in S \setminus (W_A \cup W_B)$, letting $T: = \{s''\in W_B\mid R_A[s]\subseteq R_A[s''] \}$,   some 
    $s''' \in W_A$ exists such that $T\subseteq R_B[s''']$ and, symmetrically, letting $T': = \{s'\in W_A\mid R_B[s]\subseteq R_B[s'] \}$,   some $s'''' \in W_B$ exists such that $T'\subseteq R_A[s'''']$.
\end{enumerate}
\end{remark}

In the \underline{fifth case}, the agents are not constrained by the true state, and hence neither by their response relations, and the analysis reduces to a classic game-theoretic situation, which can be analyzed similarly to the following example.  

\begin{example}
    The following table is the payoff matrix of the scenario in Example \ref{example:magical:relations}:

    \begin{center}
        \begin{tabular}{c|c|c|c}
             &$b_1$ &$b_2$ &$b_3$  \\
             \hline
             $a_1$ &$(0,0)$ &$(1,-1)$ &$(-1,1)$  \\
             \hline
             $a_2$ &$(-1,1)$ &$(1,-1)$ &$(1,-1)$  \\
             \hline
             $a_3$ &$(1,-1)$ &$(-1,1)$ &$(-1,1)$  \\
        \end{tabular}
    \end{center}

    We observe that neither $A$ nor $B$ has a dominant strategy: for instance, if $B$ plays $b_1$ then $A$ should play $a_3$, but if $B$ plays $b_2$ then $A$ should play either $a_1$ or $a_2$, and similarly, if $A$ chooses $a_1$, then $B$ must choose $b_3$, while if $A$ chooses $a_2$, then $B$ must choose $b_1$. Moreover, no outcome is a Nash equilibrium.


Let $x = (x_1, x_2, x_3)$ be the mixed strategy for Player A, and let $v$ be the value of the game. Then the linear program for Player A is:

\[
\begin{aligned}
\text{maximize } \quad & v \\
\text{subject to } \quad 
& 0x_1 - x_2 + x_3 \geq v \quad \text{(against $b_1$)} \\
& x_1 + x_2 - x_3 \geq v \quad \text{(against $b_2$)} \\
& -x_1 + x_2 - x_3 \geq v \quad \text{(against $b_3$)} \\
& x_1 + x_2 + x_3 = 1 \\
& x_1, x_2, x_3 \geq 0
\end{aligned}
\]


Let $y = (y_1, y_2, y_3)$ be the mixed strategy for Player B, and let $w$ be the value of the game. Then the dual linear program is:

\[
\begin{aligned}
\text{minimize } \quad & w \\
\text{subject to } \quad 
& 0y_1 + y_2 - y_3 \leq w \quad \text{(for $a_1$)} \\
& -y_1 + y_2 + y_3 \leq w \quad \text{(for $a_2$)} \\
& y_1 - y_2 - y_3 \leq w \quad \text{(for $a_3$)} \\
& y_1 + y_2 + y_3 = 1 \\
& y_1, y_2, y_3 \geq 0
\end{aligned}
\]


By solving the above linear programs (or checking directly), we find that the optimal strategies are:

\[
x = (0, \tfrac{1}{2}, \tfrac{1}{2}), \quad
y = (\tfrac{1}{2}, 0, \tfrac{1}{2}), \quad
v = 0
\]

That is, Player A should mix evenly between $a_2$ and $a_3$, while Player B should mix evenly between $b_1$ and $b_3$. The value of the game is zero, indicating that with optimal play, neither player gains an advantage.

    

\end{example}

The analysis above assumes that the agents are not penalized for announcing pairs of states not in $range(R)$. If they are penalized, then the payoff matrix changes and the strategies might change accordingly. 

\section{Dynamically formed Tiglath-Pileser relations}\label{Sec:Dynamically:Formed:Relations}


In this section, we discuss how the analysis of the previous section can be refined 
to formally model persuasion situations which dynamically unfold over several episodes. Such situations  
        allow us to account for the genesis of the Tiglath-Pileser relations, as they  are 
        {\em learned} by the principal, based on the their observations of what the Speaker declares. This formal analysis is applied in the two case studies of Sections \ref{sec: granpa wijnberg} and \ref{sec: king david}.


The intention is to generate a sequence $(W_n\mid n\in \mathbb{N})$ of approximations of the Speaker's winning set associated with codices which the principal extracts from the episodes of the narration. The guidance relations of these codices will  generate the sequence of approximations of the winning set   
 which  can be used to confirm or update the principal's beliefs about which claims from the narration are reliable.  


\medskip

Let us represent a story as a finite sequence $e_1,\ldots, e_k$ of episodes (each of which is a consistent set of formulas).\footnote{We assume that the narrator's story is free from internal inconsistencies that would immediately disqualify their story in the eyes of the Listener.} Let $\mathsf{Prop}$ be a non-empty set of propositional variables.  Let  $\mathsf{Prop}^V, \mathsf{Prop}^D \subseteq  \mathsf{Prop}$  such that  $\mathsf{Prop}^V \cap  \mathsf{Prop}^D = \varnothing$. Intuitively, $\mathsf{Prop}^V$ and  $\mathsf{Prop}^D$ denote the atomic facts that are, respectively, verifiable and non-verifiable (or doubtful) by the principal.\footnote{\label{ftn: across vs per episode}A more realistic assumption would have been to assign the labels `verifiable' and `doubtful' relative to each episode, i.e.~to stipulate that  $\mathsf{Prop}^V_i, \mathsf{Prop}^D_i \subseteq  \mathsf{Prop}$  such that  $\mathsf{Prop}^V_i \cap  \mathsf{Prop}^D_i = \varnothing$ for any $1\leq i\leq k$. 
However, the present stipulation allows us to assume that the sets of variables occurring in the antecedents and in the consequents of rules in codices is disjoint. 
In Footnote \ref{ftn: variable split}, we  discuss how to formally circumvent the issue presented by the occurrence of variables whose (verifiable or non-verifiable) status changes across episodes.}
Then, if $S$ denotes the set of *+ states of the world, episode $e_i$  can be associated with  the set\footnote{In what follows, we  adopt the convention of denoting all episode-bound constructions with $'$.}  
\[
E'_i = \{s\in S\mid s\models e_i\}. 
\]

Since we assume that each $e_i$ is consistent, $E'_i \neq \varnothing$. At each episode $e_i$, the principal learns the separated (but not complete and hence not pre-canonical) codex 
 
 \begin{equation}\label{eq:codex:Li'}
     L'_i :=\{\bigwedge_{y\in Y }\varphi_y\to \varphi_x  \mid Y\subseteq \mathsf{Prop}^V_i \text{ and } x\in \mathsf{Prop}^D_i \text{ and } E'_i\models \bigwedge_{y\in Y }\varphi_y\to \varphi_x  \text{ and } E'_i\models \bigwedge_{y\in Y }\varphi_y\}
 \end{equation}  
 where each $\varphi_z\in \{z, \neg z\}$ for any $z\in Y\cup\{x\}$. Since $E'_i \neq \varnothing$, it follows that $L'_i$ is consistent. In what follows we write $\varphi^\partial_x$ for $\neg x$ if $\varphi_x = x$ and $x$ if $\varphi_x = \neg x$. We allow the case when $Y = \varnothing$, in which case $\bigwedge_{y\in Y }\varphi_y\to \varphi_x$ reduces to $\varphi_x$.

We also assume the existence of a {\em background theory} $\Gamma$  for the principal/reader, which is a set of formulas, the variables of which intersect the variables of the narration but add some new ({\em background}) variables, namely the elements of $var(\Gamma) \setminus \mathsf{Prop} = \mathsf{Prop}_B$. Moreover, every formula in $\Gamma$ is of the form $\bigwedge_{x \in X} \psi_x \to \psi_y$ where $X \subseteq \mathsf{Prop}$, $y \in \mathsf{Prop}\cup \mathsf{Prop}_B$ and $\psi_z \in \{z, \neg z, \Diamond z, \Diamond \neg z\}$ for each $z \in  X \cup \{y\}$. We read $\Diamond \phi_z$ as `it is possible that $\phi_z$'.
Intuitively, the background theory encodes the general knowledge of the principal/reader which allows them to interpret the narrative.  In particular, the variables in $\mathsf{Prop}_B$ represent meta facts or judgements about the story informed by e.g.\ social norms, psychological insights, historical facts etc.  

At each episode $e_i$ we form the set 
\[
B'_i = \{p \mid p \in \mathsf{Prop}_B \text{ and } {e_i} , \Gamma \models p \} \cup \{\neg p \mid p \in \mathsf{Prop}_B
\text{ and } {e_i} , \Gamma \models \neg  p \},
\]
consisting of all literals in background variables logically entailed by the formulas describing the episode together the background theory. Intuitively, the sets $B_i'$ represent the meta-facts that the principal deduces about the story and its characters in the current episode.

So far, the sets $L'_i$ and $B_i'$ encode information at the level of each episode; let us now  define their {\em cumulative} counterparts: the sets $B_1, \ldots, B_k$ are defined inductively as follows: $B_1 = B'_1$. Assuming $B_{i-1}$ has already been defined, 
\[
B_{i} = \begin{cases}
    B_{i-1}\cup B'_i & \text{if } B_{i-1}\cup B'_i \text{ is consistent};\\
    (B_{i-1}\cup B'_i)\setminus \{\varphi_x\in B_{i-1}\cup B'_i\mid \varphi^\partial_x\in B_{i-1}\cup B'_i \} & \text{otherwise}.
\end{cases} 
\]
Intuitively, the sets $B_i$ collect together the meta-facts deduced by the principal up to that point in the story, as longs as these are consistent. These cumulative meta facts allow the reader to form conclusions about the aims of the narrator. Any inconsistencies that arise witness and flag to the principal a break in the overall narrative continuity and aims of the narrator. Note that inconsistencies at the level of the background variables are far more consequential than inter-episode inconsistencies between non-background variables which might refer to circumstances which change as the story proceeds. In contrast, inconsistencies arising in the update of the $B_i$ function as a switch between different modes of updating the cumulative codices $L_i$, inductively defined below. 

Each codex $L_i$ gives rise to a {\em winning set}  $W_i := \{ s \in S\mid s\models L_i\} = T(L_i)$, and an associated  guidance relation $\to_{L_i}$  (cf.~Definition \ref{def:guidance}). Intuitively, this guidance relation represents the principal's approximation, up to this point in the narrative, of the narrator's persuasion strategy, as formally captured by a Tiglath-Pileser relation. By Lemma \ref{lemma: T of L is T of the guidance rel}, $T(L_i) = T(\to_{L_i})$, i.e.~${\to_{L_i}}_{|W_i\times S} = \Delta_{W_i}$, i.e.~condition R2 holds for $\to_{L_i}$; moreover, since the codex is separated, every $s\in S$ is guided to some state in  $W_i$, i.e.~$\to_{L_i}^{-1}[W_i] = S$ as it is always possible to change the truth values of consequents of rules without affecting the truth values of any antecedents. In particular, condition R1 holds. In general, condition R3 would not hold for these approximations.  

Let us define $L_1, \ldots, L_k$ inductively as follows;  $L_1 := L'_1$. Assuming $L_{i-1}$ has already been defined, there are two cases to consider:

\begin{enumerate}
    \item If $B_{i-1}\cup B'_i$ is consistent, then $L_i := (L_{i-1} \cup L'_i) \setminus \mathsf{Contr}_i$ where

\[
\mathsf{Contr}_i = \left\{ \bigwedge_{y\in Y }\varphi_y\to \varphi_x  \in  L_{i - 1},   \bigwedge_{y'\in Y' }\varphi_{y'}\to \varphi^{\partial}_x \in  L'_i \: \middle| \: \{\phi_y\}_{y \in Y} \subseteq \{\phi_{y'}\}_{y' \in Y'} \textrm{ or } \{\phi_{y'}\}_{y' \in Y'} \subseteq \{\phi_y\}_{y \in Y}  \right\}.
\]

\item As mentioned above, if $B_{i-1}\cup B'_i$ is inconsistent, then this signals to the reader an apparent anomaly in the aims of the writer, insofar as the reader has inferred them up to this point. 
This leads the reader to conclude that the episode $e_i$ does not provide information that could be used to update the current approximation of the writer's persuasion strategy, as represented by the guidance relation associated with codex $L_{i-1}$, nor the current approximation of their winning set $W_{i-1}$.  Therefore, we set 
$L_{i} := L_{i-1}$.  

\end{enumerate}

This allows us to define a sequence of sets of formulas $T'_1, \ldots, T'_k$, representing what the reader considers reliable information related to each episode. 
        %
         %

\[
T'_i = \begin{cases}
    \{\varphi_x \mid x \in \mathsf{Prop}_i^V\cup \mathsf{Prop}_i^D \text{ and } E'_i\models \varphi_x \}  & \text{if } B_{i-1}\cup B_i' \text{ inconsistent and } E'_i \cap W_{i-1} = \varnothing \\
    \{\varphi_x, \Diamond \varphi_y, \Diamond \varphi^{\partial}_y \mid x \in \mathsf{Prop}_i^V, y\in \mathsf{Prop}_i^D, E'_i\models \varphi_x, E'_i\models  \varphi_y \}  &\textrm{otherwise}
    \end{cases}
\]

Informally, if $B_{i-1}\cup B_i'$ is inconsistent and  $E'_i \cap W_{i-1} = \varnothing$, then in episode $e_i$ the writer's claims go against the current approximation $W_{i-1}$ of their winning set and overarching aims as surmised by the principal up to this point in the narration and captured in $B_{i-1}$. Under the assumption that the narrator's persuasion strategy is captured by a Tiglath-Pileser relation (i.e.\ satisfying R3), the principal is led to conclude that the narrator has been forced to truthfully declare something outside their winning set. In other words, the principal can deduce that the epistemic constraints under which the writer operates make it impossible for them to pursue their aims irrespective of the truth in the present episode; this allows the principal to regard  as reliable information {\em also} the unverified claims at this point in the narration which fall outside the writer's winning set.

Now, analogously to the sets $B_1, \ldots, B_k$, we define the sets $T_1, \ldots, T_k$, each of which representing what the reader considers reliable information up to that point in the narration. Preliminarily,   for any set $\{\psi_x \mid x \in X \}$, we define the \emph{direct consequences of $\{\psi_x \mid x \in X \}$ via $\Gamma$} to be
\[
DC_\Gamma(\{\psi_x \mid x \in X \}) := \{\psi_y \mid \bigwedge_{x \in X'} \psi_x \to \psi_y \in \Gamma \text{ and } \{\psi_x \mid x \in X' \}\subseteq \{\psi_x \mid x \in X \} \}.
\]
Let $T_1 := T'_1$. Assuming $T_{i-1}$ has already been defined,

\begin{align*}
    T_i &=  ((T_{i-1}\cup T_i')\cup DC_\Gamma(T_{i-1}\cup T'_i)) \setminus \\
    &(\{\phi_x\in DC_\Gamma(T_{i-1}\cup T'_i) \mid \phi^\partial_x\in DC_\Gamma(T_{i-1}\cup T'_i) \}\cup \{\Diamond\phi_x, \Diamond \phi^\partial_x\mid \varphi_x\in DC_\Gamma(T_{i-1}\cup T'_i) \text{ or } \varphi^\partial_x\in DC_\Gamma(T_{i-1}\cup T'_i) \})
\end{align*}

Note that, as suggested by the notation in the above definition, the only possible sources of inconsistency are the propositional literals, since formulas of the form $\Diamond \phi_x$ never occur negated in the elements of $\Gamma$. Hence, intuitively, the reliable information acquired by the reader up to episode $e_i$ consists of the direct consequences acquired up to $e_i$ where logical inconsistencies are removed, together with  `doubts',  whenever `doubts' are superseded by reliable factual information.

\medskip

The procedure described above accounts for the formal representation of the `writer's spin' as understood by the principal/reader, in the form  of the Tiglath-Pileser relation (i.e.~the guidance relation associated with the codices gauged by the principal up to each episode in the narration), as well as the formation of the reader's beliefs as to which pieces of information in the narration are  reliable.  However, note that  the definition of the codices $L_i$ leaves the reader vulnerable to the possibility that a strategic writer who understands this procedure could engineer the narration so as to manipulate  the listener into believing unverified information. Therefore, a reader who is aware of this possibility might update the information they consider reliable in a different way, which we present below.

\paragraph{Update protocol for the strategic listener/reader.}

From the discussion above, it emerges that the way in which  the listener can be manipulated into believing unverified information involves creating a series of episodes which first allow the reader to build some fundamental convictions about the aims of the narrative, only to create a narrative fracture (the fracture point being the episode at which $B_{i-1}\cup B_i'$ is inconsistent) in which what is claimed is at odds with the general aims of the narrator as they are understood by the reader up to that point (as represented by $E_i'\cap W_{i-1} = \varnothing$), and  the reader's  fundamental convictions formed up to that point are contradicted. Engineering such a moment is {\em expensive}, as it not only requires significantly greater complications in weaving the story, but also has detrimental effects on achieving some of the ostensible narrative goals, e.g.~concerning the reputation of some of the characters in the story, and revealing certain details about which the writer would be reasonably expected to rather keep quiet.   
As discussed above, the ordinary reader might be tempted to apply Ockham's razor and  assume that, if the writer got to  a point in the story in which he had to make claims which are detrimental to his aims, then these claims could not have been avoided, softened or `given a spin' (and this is why they can be trusted). However, the strategic reader might wonder whether the significant costs incurred by the writer could actually {\em be worth their while}; that is, whether the benefits of  being able to plant certain {\em unverifiable} pieces of information in the mind of the reader would justify the  costs of this expensive maneuver. If these costs are justified, then the reader should better believe {\em the opposite} of what the writer claims concerning these variables. This is captured in the following, alternative update rule for $T'_i$:


\[
T'_i = \begin{cases}
    \{\varphi_x, \varphi^\partial_y \mid x \in \mathsf{Prop}_i^V, y\in \mathsf{Prop}_i^D , E'_i\models \varphi_x, E'_i\models  \varphi_y\}  & \text{if } B_{i-1}\cup B_i' \text{ inconsistent and } E'_i \cap W_{i-1} = \varnothing \\
    \{\varphi_x, \Diamond \varphi_y, \Diamond \varphi^{\partial}_y \mid x \in \mathsf{Prop}_i^V, y\in \mathsf{Prop}_i^D, E'_i\models \varphi_x, E'_i\models  \varphi_y  \}  &\textrm{otherwise}
    \end{cases}
\]
At this juncture, a more sophisticated update protocol might handle different doubtful variables differently (e.g., negate, prefix with modalities, or remove them) depending on their relevance to the context of the episode. This will not be pursued here.

\section{Modelling the story of the small businessman and the banker}
\label{sec: granpa wijnberg}
The present section focuses on a case study from \cite{conradie2019probabilistic}: 
\begin{quote} Around 1950, there was a small businessman in Amsterdam whose main business was to sell the products of foreign textile manufacturers to Dutch clothing firms. Like most small businessmen in
Amsterdam at the time, he banked with the Amsterdamsche Bank (which later became the present ABN AMRO).
One day, he received an invitation to lunch with one of the directors of that bank. This invitation puzzled him a great deal, because he did not know this director personally, and a small businessman like him usually only dealt with bank employees at much lower levels. However, he accepted the
invitation and showed up for the lunch at the top floor of the bank’s headquarters in the city centre.
During the copious lunch, the bank director talked about all kinds of general subjects and asked the businessman's opinion about the economic climate in Amsterdam. Rather than being flattered, he  found it hard to imagine that he was invited to provide opinions about matters the bank knew better than he.
When the dessert was served, the banker mentioned aside some other matter the name of a certain
Amsterdam firm, which was an important client of the businessman. This firm, the bank director said, was doing
very well under the present solid leadership.
The small businessman realized that this must have been the point of the whole lunch. And if this large bank went to so much effort to increase the confidence of one small businessman in this
firm, it must have been very important to the bank that the businessman believed that the firm was doing well.

The small businessman said he wanted to wash his hands, although coffee still needed to be
served, but instead of walking to the bathroom he ran down the stairs and onto the street to find a
telephone booth and call the office to stop all deliveries to that firm and also claim back any supplies that
had already been delivered.
Two weeks later, that firm went bankrupt and it turned out that the bank not only was its major creditor
but also had preferential right to sell off any stocks in the possession of the firm to pay back the debt to
the bank before other creditors would be satisfied.
\end{quote}


In the following subsections, we present static and dynamic formal analyses of this case study  as a persuasion scenario in which the banker is the Speaker $A$ and  the businessman is the Listener $B$. 

\subsection{Static analysis}
Our account of the narration includes the following variables:

\begin{tabular}{c|c|l}
    D &$k$ &bank is {\bf k}nowledgeable about current business conditions\\
    V &$q$ & the bank asks  {\bf q}uestions about the economic climate\\
    V &$p$ & the bank  solves its clients' {\bf  p}roblems\\
    B & $n$ & individual clients {\bf n}eed the bank more than the bank needs each of them\\
    D & $f$ & the {\bf f}irm is solid
\end{tabular}

\noindent The table above also reports the verifiable/doubtful/background status of each variable, in preparation for the discussion of Section \ref{ssec: dynamic analysis grandpa}. The average businessman's background theory is given by:
\[
\Gamma = \{k \wedge p \to n, q\to \neg n \}.
\] 
In other words, if the bank is knowledgeable about the current business conditions and solves its clients' problems then individual clients need the bank more than the bank needs each of them, and if   the bank asks  questions about the economic climate then it is not the case that individual clients need it more than it needs each of them. 
Let $S=2^{\mathsf{Prop}}$ be the set of profiles for $\mathsf{Prop}=\{k,q,p,n, f\}$. Let $s_0=(1,1,1,1,0)$ be the true state. 
The interaction between an ordinary businessman and the banker corresponds to case 3 of the hierarchy  discussed in Section \ref{ssec:models of persuasion via TP relations}.
That is, the businessman $B$ is strategic but believes that the banker $A$ is non-strategic, while   $A$ is strategic, believes that $B$ is strategic and that $B$ believes  $A$ to be non-strategic. The  ordinary  businessman  believes that the bank  tries to appear knowledgeable whenever they can. This leads them
 to believe that the speaker's winning set is  $\val{k} = \{s\in S\mid s(k) = 1\}$ and 
 that the speaker's guidance relation $R_A \subseteq S \times S$ is
such that for any $s \in S$, 

\begin{enumerate}
    \item if $s(k) = 1$, then $R_A[s]=\{s\}$,
    \item if $s(k) = 0$ and   $s'(y) =s(y)$ for all $y\in \mathsf{Prop}\setminus\{k\}$ and $s'(k) = 1$ and $Th(s')\cup \Gamma$  is consistent, 
    then $R_A[s]=\{s'\}$,
    \item if $s(k) = 0$ and  $s'\not\models \Gamma$, where $s'$ is the $k$-variant of $s$ s.t.~$s'(k) =1$, then $R_A[s]=\{s\}$.
\end{enumerate}

Note that $R_A \in \mathsf{TP}(S,\val{k})$. 
In this particular case, the speaker claims $\neg k$, $p$, $q$, $f$. This is formalized as 
the  speaker  announcing the set of states $S'=\val{\neg k \wedge q \wedge p \wedge f} =\{(0,1,1,0,1), (0,1,1,1,1) \}$. As per the discussion in Section \ref{ssec:models of persuasion via TP relations}, the ordinary businessman guesses that the true state is in $R_A^{-1}[S']=S' \subseteq \val{f}$. Hence, the banker succeeds in persuading the ordinary businessman that the firm is solid.

Unlike the ordinary businessman, the small businessman described in the story above  believes that the banker is strategic. So, as in  case 4 of Section \ref{ssec:models of persuasion via TP relations},  he cannot draw any conclusions regarding the true state without making further behavioural assumptions about the banker.  The small businessman notices that, in this case,  $R_A^{-1}[S'] \cap W_A =\varnothing$; that is, the banker's claims are inconsistent with the businessman's assumptions about the banker's winning set. However, since the businessman assumes that the banker is strategic, they do not come to same conclusion as the ordinary businessman, and look for a different explanation. Among the claims made by the banker, the only unverifiable one concerns the solidity of the firm. Hence, the businessman concludes that the banker's actual winning set must have been $\val{f}$. However, this static framework does not allow for making updates about the winning set  of the speaker or their associated Tiglath-Pileser relations.

\subsection{Dynamic analysis}
\label{ssec: dynamic analysis grandpa}

For the sake of the dynamic analysis, we define the following two episodes, $e_1$ and $e_2$:
\begin{enumerate}
    \item[$e_1$] previous relationship with the bank (the bank solves problems and sets the terms)
    \item[$e_2$] lunch    
\end{enumerate}

In the successive episodes the following propositions are asserted:

\begin{tabular}{c|c}
    $e_1$ &  $k\wedge p$\\
    $e_2$ &  $\neg k \wedge q \wedge f$
\end{tabular}

The codices learned from each of the episodes are:
\[
L'_1 = \{k, p \to k \} \quad \quad 
L'_2 = \{\neg k, f, q \to \neg k, q \to f\}
\]
Notice that, as we allow for the conjunction in the premise of each rule to be empty, the sets above include also the doubtful literals satisfied in the given episode. The sets $B'_i$ for $i = 1, 2$ are computed as follows:
\[
B'_1 = \{n \} \quad \quad 
B'_2 = \{\neg n\}
\]
Since $B_1\cup B_2' = \{n, \neg n\}$ is inconsistent, the sets $B_i$ and $L_i$ are computed as follows: 
\[
    B_1 = \{n \} \quad \quad
    B_2 = \varnothing
\quad \quad
    L_1 = \{k, p \to k \} \quad \quad
    L_2 = L_1
\]
From the codex $L_1$, the Listener extracts the approximation $W_1 = \val{k} =  \{s\in 2^{\mathsf{Prop}}\mid s( k) = 1\}$ of the Speaker's winning set. 
 The set $T_1'$ consists of the verifiable literal $p$ which is true at $e_1$, together with the assertions of the possibility of the doubtful literals $k$ and $\neg k$.

 \[
    T'_1 = \{p, \Diamond k, \Diamond \neg k\} 
\]
 
In episode $e_2$, the conditions indicating an epistemic fracture  are verified: indeed, in addition to $B_1\cup B_2' = \{n, \neg n\}$ being inconsistent, in episode $e_2$, the banker makes declarations the truth-set of which is disjoint from $W_1$. These two conditions trigger an  extraordinary update of the set $T_2'$, which can be performed in two possible ways:

\[    
    T_2' = \{q, \neg k, f \} \quad\quad T'_2 = \{q,  k, \neg f\} 
\]
The set $T'_2$ above on the left reflects the update of a non-strategic Listener, as expected by the banker in the story; this Listener detects the epistemic fracture and believes that the banker's claims in $e_2$ can be trusted precisely because they contradict the image that the bank had projected up to that point.  In contrast, the $T'_2$ above on the right reflects the update of a strategic Listener, who recognizes the costs (in terms of time, effort, and projected image) involved in generating the epistemic fracture, and understands that  the stakes involved in making the Listener believe the unverified claims must be  high enough  to justify this investment. While $T_1 := T_1'$, the two versions of $T_2'$ above give rise to the following two versions of the set $T_2$, respectively:  
\[   
    T_2 = \{p, \neg k, q, f \} \quad\quad T_2 = \{p,  k, q, \neg f\}
\]
These two sets have been obtained by taking the union of $T_1$ and $T_2'$, adding their direct consequences, and removing the doubts whenever they are superseded by reliable information; however, the two different updates of $T_2'$ make a crucial difference: by $e_2$, the conclusion that the firm is solid has been planted in the mind of the non-strategic businessman, while the strategic businessman has concluded the exact opposite.

\section{Modelling Baruch Halpern's account of King David's story}
\label{sec: king david}

We now turn our attention to the analysis of Baruch Halpern's account of the King David story, as already discussed in the introduction. In Section \ref{ssec:Halpern:informal}, we provide an informal account of the main episodes of this story, together with Halpern's analysis. This is formalized in Section \ref{ssec: modelling Halpern} using the framework introduced in Section \ref{Sec:Dynamically:Formed:Relations}.

\subsection{The informal analysis}\label{ssec:Halpern:informal}

In the present section, we summarize some of the main episodes of King David's story as related in 1 Samuel, 2 Samuel, and the opening of 1 Kings,  in the light  of Baruch Halpern's analysis in \cite{halpern2003david}. 

Ancient royal historiography is often imagined as straightforward propaganda: kings win, enemies fall, disasters disappear from the record. The classic model according to Halpern is the Assyrian court chronicle, the sort of inscription associated with rulers like Tiglath-Pileser III, which can be interpreted
through the Tiglath-Pileser principle, discussed in the introduction. The governing rule of these chronicles is simple: if something happened and it was not a disaster, narrate it as a resounding triumph; if it was a disaster, pass over it in silence as far as possible without completely compromising the overall credibility of the report given what the purported original audience knows. The result is what we may term  the ``Nineveh Times'' style of news-reporting:   observable \footnote{That is, observable from the perspective of the intended audience, e.g., the population at large.} facts are reported, but selectively, and always favourably. These observables might be embellished with unobservables, e.g.~claims of overwhelming victory or divine favour. In this mode of writing,  readers are trained in a negative deduction procedure,  
which allows them to infer or approximate the
truth both from the `spin' with which the writer seems to present the narration and  from the writer's omissions. Since the audience is aware that unfavorable news is not recorded, the propagandistic effect of the narration is weakened, as readers of the ``Nineveh Times'' may wonder and worry about what they are not told. 

The narrative of David in 1 Samuel, 2 Samuel, and the opening of 1 Kings  adopts a 
more sophisticated strategy. At first glance, it appears to abandon the ``Nineveh Times'' style of news-reporting, in so far as
it includes potentially unfavourable information the ``Nineveh Times'' might have
preferred to omit: 
civil war, court intrigue, dynastic rebellion, adultery, arranged death are displayed in broad daylight. The historian seems almost reckless in his willingness to expose the king to suspicion.
What we may call the ``Jerusalem Times'' style of  news-reporting operates differently. It creates the impression that all newsworthy observables are reported, including those that could damage the royal image: the narrator not only acknowledges these facts, but also {\em supplies additional (often unverifiable) details}, especially at precisely those points where the reader might expect less than full disclosure. 
Because the narrator demonstrably anticipates and flags potential suspicions, the reader feels no need to search for hidden scandals beyond those identified in the text. Suspicion is not erased, it is managed.

This creates a new dynamic. Unlike the Assyrian reader, who discounts good news and assumes suppression of bad, the reader of Samuel is trained to believe that the unpleasant facts have already been disclosed. The narrator flatters the reader’s intelligence by openly wrestling with embarrassing material and by going to elaborate lengths to explain, contextualize, or morally frame it. Even when the explanations are not wholly convincing, the gesture itself constrains the field of suspicion, as the reader’s  doubt is thereby focused on the issues the narrator signals, and nowhere else. 

The episode of Bathsheba will prove pivotal in this more sophisticated narrative strategy. Up to this point, the writer has portrayed David in  as positive a light  as possible. 
Although the narrator has described many events which  could have 
put the king in an unfavourable light, he has gone to great lengths to explain away anything in the
narrative that could have suggested that the king was at fault; for instance, when an enemy dies in suspicious circumstances, it is explained that  David had not ordered his death and was very distraught about it.  However, now it appears that the writer gratuitously sacrifices this carefully built image, telling a  story that he could have narrated far more favourably. But, according to Halpern \cite{halpern2003david}, once the larger narrative strategy is understood, this “sacrifice” can be seen as functional: it reinforces the impression of total candour, and closes off more dangerous lines of doubt. The writer  does not suppress incriminating observables, and he even adds seemingly ``gratuitous'' unobservables, such as the secret letter from David to Joab. By doing so, he creates a far more resilient form of royal apology than any Assyrian victory inscription could achieve.

What follows, then, is not just a retelling of key episodes in David’s life, but an exploration of how their narration differs from the Assyrian model, and how this difference calls for a more sophisticated application of the Tiglath-Pileser principle, leading to a new understanding of the
narrator's real agenda, which may be less about  glorifying David, and more about  strengthening the legitimacy of the House
of David as a royal lineage, especially from Solomon onwards. 



\paragraph{Episode 1.} David first appears in the narrative when the prophet Samuel anoints him in secret while Saul still reigns (1 Samuel 16). The act is quiet but politically explosive: the divine spirit departs from Saul and comes upon David. From that moment, tension is unavoidable. The narrative does not conceal the enmity that follows. Saul’s jealousy, his attempts on David’s life, David’s life as a fugitive—these are all laid out in full view. A court chronicle in the ``Nineveh Times'' style would likely have erased such a dynastic fracture. Here, however, the reader sees it all. Yet at every critical juncture David is shown refusing to harm “the LORD’s anointed.” Twice he spares Saul when he could have killed him (1 Samuel 24, 26). The embarrassing reality of civil hostility is fully exposed, but the interpretation is steered: David is not an opportunist; he is the reluctant heir, waiting for God’s timing. Because the text does not hide the tension, its exculpation of David feels earned.

\paragraph{Episode 2.} The episode of Nabal (1 Samuel 25) sharpens this pattern. After  protecting the  flocks of the wealthy  farmer Nabal, David demands a payoff. When Nabal refuses, David leads a detachment of his men to massacre Nabal and his people. The narrative does not hide the basic structure of a protection racket David is organizing. Violence is only prevented when Nabal's wife Abigail intercepts David with gifts of food and drink. After her return, Nabal dies  struck down by God, and David marries his widow Abigail. On the face of it, this narrative could have been troubling: a small-scale gang leader threatens a wealthy landowner who soon conveniently dies. Yet, the story’s willingness to expose David’s  protection racket works as a token of trust, which---even if the reader was suspicious of David's involvement in Nabal's death
---builds confidence that nothing worse is being hidden.

\paragraph{Episode 3.} When Saul and his three sons die in a battle against the Philistines (1 Samuel 31, 1 Chronicles 10), the political beneficiary is obviously David. The text does not minimize this, nor does it hide the fact that, at the time of this battle, David and his warriors were under the protection and in the service of  the Philistine king Achish. However, the narrator provides an elaborate storyline that explains why David and his forces were not there on the day of the battle. Three days later, an Amalekite from the Israelite army camp brings David news of the defeat and death of Saul and his sons, and claims to have delivered the final blow to Saul (2 Samuel 1). David executes him for daring to strike the LORD’s anointed, and composes a lament of extraordinary beauty over Saul and Jonathan. Again, the suspicious circumstance that David profited from his rival’s death is fully visible. The narrative counters, not by denying the advantage, but by emphasizing David’s grief and his principled refusal to legitimize regicide. The narrator expends considerable effort here, almost ostentatiously, to distance David from any whiff of complicity. In Halpern’s terms, this is one of the early “sacrifices” of narrative energy: the narrator works hard to defend the king where suspicion is plausible. This narrative strategy habituates  the skeptical reader to expect David to be elaborately exculpated whenever the facts allow for it. Then comes the Bathsheba episode.

\paragraph{Episode 4.} In 2 Samuel 11, while David's army is laying siege to the capital of the Ammonites,  David seduces Bathsheba, wife of Uriah the Hittite, an elite soldier in David's army. Bathsheba falls pregnant and David attempts to cover up his adultery  by summoning Uriah home from the front.  The narrative lingers over the details: Uriah refuses to go down to his own house and sleep with his wife; he sleeps at the palace entrance; David sends him back to battle carrying the letter to Joab in which he orders Joab to organize an attack on the city walls where Uriah is likely to be
killed. 
A host of incriminating unobservables (first and foremost the secret letter to Joab) are placed in full view of the reader.

Up to this point in the larger story, the reader has been trained to expect that such dangers to David's reputation will be offset. But here no attempt at  exculpation is made; in fact, the story emphasises David's failings and punishment: the prophet Nathan confronts David (1 Samuel 12) and prophesies violence against David’s house. The first child born to David and Bathsheba dies after the confrontation with Nathan. Only after all this is over do they conceive
Solomon, who in this way is born at many removes from ``sin". 

This is the second and far more dramatic ``ostensible sacrifice''. The narrative that has hitherto so carefully shielded David now exposes him. Because the narrator has demonstrated his willingness to tell the worst possible story about a king betraying a loyal officer in his own army, the reader is disinclined to suspect that something even worse lies concealed as, apparently, the worst has already been confessed.

Here, Halpern asks a pointed question. The narrator did not need to tell the story this way. He could have portrayed Uriah as a loyal hero who fell bravely in battle, and David as compassionately marrying his widow. Instead, he chose to blacken David’s character with adultery and murder. Why make such a sacrifice?

Halpern’s suggestion is that this ``ostensible sacrifice'' of David’s image serves a larger  purpose. By narrating the sin and its punishment, the historian seals off a  more dangerous suspicion; namely, that Solomon might actually be Uriah’s son. 
The narrator's trustworthiness seems to be guaranteed precisely by his display of this plethora of damning unobservable details without any attempt to place them in a better light. 
The sequence is emphatic: conception in sin, exposure, judgment, death of the first son; then a second son, Solomon, is born after the reckoning. This creates a narrative buffer to protect Solomon's paternity from doubt. 
Because the narrator was willing to humiliate David, the reader is relieved of having  to imagine   worse scenarios.


\paragraph{Episode 5.} The consequences of David's sin unfold in the tragedy of Absalom. A fraught relationship between David and Absalom (1 Samuel 13 - 14) culminates in Absalom rebelling and usurping the throne (1 Samuel 15). Absalom's popularity is not hidden by the narration: the Israelites side with him, and David has to flee Jerusalem in humiliation. 
David regains the throne only thanks to his professional army of foreign mercenaries. 
The revolt nearly tears the kingdom apart. 
When Absalom is killed by Joab despite David’s express and public command not to harm him, the king’s grief overwhelms him. Here we see the old pattern of exculpation resumed and reinforced in the mind of the reader: David clearly benefits from Absalom's death, but the details of the narrative are arranged to clear him of any complicity.  The civil war is not minimized, but it is framed as the  consequences of David’s own sin, in accordance with Nathan’s prophecy. Hence, the narrative frames civil war as a punishment for moral turpitude; 
the instability of David's reign is acknowledged, yet contained within a moral explanation. 

\paragraph{Episode 6.} In 1 Kings 1, the reader finds king David very advanced in years and in failing health. His son Adonijah emerges as his natural heir, as the narrative makes clear: he is David’s son, older than Solomon, and his public celebrations and broad support (also by Joab, the head of the army) suggest that he is precisely the successor everyone would expect. 
%
However, Bathsheba and the prophet Nathan conspire to extract from the apparently senile David a declaration in favour of Solomon's succession and instructions to have him publicly anointed as king. This is carried out, and Adonijah, Joab, and their followers are killed. 


Here the earlier Bathsheba narrative does its quiet work. The succession struggle is framed as a contest between two sons of the same king. Because the Bathsheba episode has already exhausted the reader’s suspicion by exposing adultery, conspiracy, and punishment, the possibility that Solomon might be the spearhead of a coup organized by David's personal guard does not rise on the reader’s horizon. The dark story that seemed so damaging to David has, paradoxically, made Solomon’s legitimacy narratively secure.

In this light, the Bathsheba episode is not gratuitous confession, but rather, strategic genius. The narrator sacrifices David’s moral image in order to fortify the dynasty’s 
continuity. The reader, having seen the apparent worst, feels no need to search for something worse still,
%
%
%
and that is precisely why the most dangerous question of all—the question of Solomon’s paternity—never quite occurs.

\subsection{The  formalization}
\label{ssec: modelling Halpern}

In line with the discussion of Section \ref{ssec:Halpern:informal}, we consider the following episodes:

\begin{enumerate}
    \item[$e_1$] Samuel anoints David. 
    \item[$e_2$] Death of Nabal.
    \item[$e_3$] Death of Saul and Jonathan.
    \item[$e_4$] Death of Uriah.
    \item[$e_5$] Death of Absalom. 
    \item[$e_6$] Succession of Solomon. 
\end{enumerate}

Our account of the narration includes the following variables:\footnote{\label{ftn: variable split} By their intended meaning, the variables $o_1$ and $o_2$ are such that $o_2\to \neg o_1$; hence, alternatively, we could have replaced both variables with a single variable; however, the status of this variable would have to change across episodes, hence to avoid the extra technical overhead that  this would involve (cf.~Footnote \ref{ftn: across vs per episode}), we introduced two separate variables.}

\begin{tabular}{c|c|l}
    V &$a$ &death {\bf a}dvantageous to David occur\\
    D &$a'$ & David is {\bf a}nointed\\
    D &$a''$ &David harbours {\bf a}nimosity against the deceased\\
    D &$i$ &David is {\bf i}nvolved in a murder\\
    V &$e$ &David {\bf e}ulogises the deceased\\
    D &$o_1$ &the {\bf o}bstacle to David's aims dies naturally\\
    V &$o_2$ &the {\bf o}bstacle to David's aims dies violently\\
    
    D &$s$ &Solomon is the biological {\bf s}on of the reigning king \\
    D &$h$ &Solomon is David's designated {\bf h}eir\\
    D &$g$ &David is anointed by {\bf G}od as king\\
    V &$p$ &The {\bf p}eople support Adonijah\\
    V &$j$ &{\bf J}oab supports Adonijah\\
    V &$k$ &Adonijah publicly celebrates his {\bf k}ingship\\
    V &$t$ &Solomon ascends to the {\bf t}hrone\\
    V &$m$ &Solomon {\bf m}urders everyone connected to Adonijah\\
    D &$c$ &Adonijah attempts a {\bf c}oup\\
    B &$f$ &Solomon stages a foreign coup\\
    D &$d$ &David {\bf d}enies Adonijah's claim to the throne \\
    B &$b$ &David behaves in a way {\bf b}efitting a king \\
\end{tabular}

\medskip
\noindent where the labels V, D, B stand for `verifiable', `doubtful', `background', respectively. The literals corresponding to these variables will be evaluated in the successive episodes of the narrative, but the ultimate intention is to interpret them over the narrative as a whole. Therefore if a literal is concluded at the end of the narrative, we read that as meaning that it held stably true in the last stretch of the narration, which could include that it held throughout.

In the successive episodes the following propositions are asserted:\\
\\
\begin{tabular}{c|c}
    $e_1$ &  $g\wedge a'$\\
    $e_2$ & $a \wedge a'' \wedge \neg i \wedge \neg e \wedge o_1$ \\
    $e_3$ &$a \wedge \neg a'' \wedge \neg i \wedge e \wedge o_2$  \\
    $e_4$ &$a \wedge  i \wedge \neg e \wedge o_2  \wedge s$  \\
    $e_5$ &$a \wedge \neg a'' \wedge \neg i \wedge e \wedge o_2$  \\
    $e_6$ &$h \wedge p \wedge j \wedge k \wedge t \wedge m \wedge c \wedge \neg d$
\end{tabular}

The set of profiles is therefore $S = 2^{\mathsf{Prop}'}$, where $\mathsf{Prop}'$ is the set of variables in the table above, i.e., $\mathsf{Prop}' = \mathsf{Prop} \cup \mathsf{Prop}_V \cup \mathsf{Prop}_D$. 

As described above, we associate  the following sequence of sets of profiles with the sequence of episodes:

\begin{align*}
    E'_1 &= \{s \in S  \mid s(g) = 1, s(a') = 1\}\\
    E'_2 &= \{s \in S  \mid s(a)=1, s(a'') = 1, s(i) = 0, s(e)= 0, s(o_1)=1\}\\
    E'_3 &= \{s \in S  \mid s(a)=1, s(a'') = 0, s(i) = 0, s(e)= 1, s(o_2)=1\}\\
    E'_4 &= \{s \in S  \mid s(a)=1,  s(i) = 1, s(e)= 0, s(o_2)=1,  s(s)=1\}\\
    E'_5 &= \{s \in S  \mid s(a)=1, s(a'') = 0, s(i) = 0, s(e)= 1, s(o_2)=1\}\\
    E'_6 &= \{s \in S  \mid  s(h) = s(p) = s(j) = s(k) = s(t) = s(m) = s(c) = 1, s(d) =0\}
\end{align*}

From this the following episode-related codices are generated: 
    %
    %
    %
%

\begin{align*}
    L'_1 &= \{ g, a' \}\\
    L'_2 &= \{u, a \to u, \neg e \to u, a \wedge \neg e \to u \mid u \in \{a'', \neg i,  o_1 \} \}\\
    L'_3  &= \{v,  a \to v, e \to v, a \wedge e \to v, \mid v \in \{\neg a'', \neg i,  o_2 \} \}\\
    L'_4 &= \{\bigwedge Y \to x    \mid Y \subseteq \{ a, \neg e, o_2 \}, x \in \{i, s \} \}\\
    L'_5 &= L'_3\\
    L'_6 &= \{\bigwedge W \to v' \mid W\subseteq \{p, j, k, t, m\} \text{ and } v' \in \{h, c, \neg d  \} \}
\end{align*}

We assume that the background theory of the reader is given by 
\[
\Gamma = \{ o_2\to \neg o_1, g\wedge a'\to b, a \wedge i \to \neg b, \Diamond\neg s \wedge \Diamond\neg h \to \Diamond f,  \neg s \wedge \Diamond\neg h \to \Diamond f  \}
\]
We therefore get $B'_1 = \{b\}$, $B'_2 = B'_3 = B'_5 = B'_6 = \varnothing$, $B'_4 = \{ \neg b\}$. And hence $B_1 = B_2 = B_3 = \{ b\}$, then $B_4 = B_5 = B_6 =  \varnothing$.

We are now ready to calculate the codices $L_i$: 
\begin{align*}
    L_1 &= \{ g, a' \}\\
    L_2 &= \{ g, a' u, a \to u, \neg e \to u, a \wedge \neg e \to u \mid u \in \{a'', \neg i,  o_1 \} \}\\
    %
    %
    L_3  &= \{g, a' \neg i, o_1, o_2 \} \cup \{ a \to u' \mid u' \in \{\neg i, o_1, o_2 \}\}\\
    &\cup  \{ y' \to u \mid y'\in \{e, e \wedge a\}, u \in \{a'', \neg i,  o_1 \}\} \cup \{ y \to v  \mid y \in \{\neg e, \neg e \wedge a\}, v \in \{ \neg a'', \neg i,  o_2 \} \}\\
    L_4  &= L_3\\
    L_5 &= L_4 \cup L'_5 = L_3 \cup L'_3 = L_3\\
    L_6 &= L_5 \cup L'_6.
\end{align*}

\begin{remark}
The two cases in the definition of the codex $L_i$ in Section \ref{Sec:Dynamically:Formed:Relations} are illustrated by
the computations of $L_3$ and $L_4$ above; indeed, since $B_2\cup B'_3 = \{b\}$ is consistent, the codex $L_3$ is defined as the union of $L_2$ and $ L'_3$ to which the inconsistencies, if there are any, have been removed. In this case, the inconsistencies which need to be removed originate from  proposition $a''\in\mathsf{Prop}_D$, which is true in  $e_2$ and false in  $e_3$. Consequently, not only $a''$ and $\neg a''$, but  also $a\to a''$ and $a\to \neg a''$,  have  been removed from the ensuing codex $L_3$. Moreover, since $B_3\cup B'_4 = \{b, \neg b\}$ is inconsistent, $L_4'$ is not used to  update  $L_3$, which remains unchanged.
\end{remark}
From the codices above, we can derive their corresponding  winning sets $W_i: = T(L_i)$.

Let us compute the sets $T_i'$. $T'_1 = \{\Diamond g, \Diamond \neg g, \Diamond a', \Diamond \neg a' \}$, since there are no verifiable variables and there are no inconsistencies in $B_1$;
$T'_2 = \{a, \neg e, \Diamond i, \Diamond \neg i, \Diamond a'', \Diamond \neg a'', \Diamond o_1, \Diamond \neg o_1\}$; $T'_3 = \{a,  e, o_2, \Diamond a'', \Diamond \neg a'', \Diamond i, \Diamond \neg i \}$;
the interesting one is 
$T'_4 = \{a, i, \neg e, o_2,  s \}$; notice that $i,  s$ are  `non verifiable' literals; however, the reader considers their truth reliable since $W_3\subseteq \{s\in S\mid s\models \neg i\}$ and $E_4\subseteq \{s\in S\mid s\models i\}$, hence $W_3\cap E_4 = \varnothing$, and moreover $B_3\cup B'_4 = \{b, \neg b\}$ is inconsistent. Then $T'_5 = \{a,  e, o_2, \Diamond a'', \Diamond \neg a'', \Diamond i, \Diamond \neg i \}$ and $T'_6 = \{p,j,k,t,m, \Diamond c, \Diamond \neg c, \Diamond h, \Diamond \neg h, \Diamond d, \Diamond \neg d\}$.

We now calculate the sets $T_i$ as follows: 

\begin{align*}
    T_1 &=  \{\Diamond g, \Diamond \neg g, \Diamond a', \Diamond \neg a' \}\\
    T_2 &= \{a, \neg e, \Diamond a'', \Diamond \neg a'', \Diamond i, \Diamond \neg i, \Diamond o_1, \Diamond \neg o_1, \Diamond g, \Diamond \neg g, \Diamond a', \Diamond \neg a'\}\\
    T_3 &= \{a,  o_2, \neg o_1,\Diamond a'', \Diamond \neg a'',  \Diamond i, \Diamond \neg i,   \Diamond g, \Diamond \neg g, \Diamond a', \Diamond \neg a'\}\\
    T_4 &=  \{a,  \neg e, i, o_2, \neg o_1,  s, \neg b, \Diamond a'', \Diamond \neg a'', \Diamond g, \Diamond \neg g, \Diamond a', \Diamond \neg a' \}\\
    T_5 &= \{a, i, o_2, \neg o_1,  s, \neg b, \Diamond a'', \Diamond \neg a'', \Diamond g, \Diamond \neg g, \Diamond a', \Diamond \neg a' \}\\
    T_6 &= \{a,  i, o_2, \neg o_1,  s, \neg b,  p,j,k,t,m, \Diamond a'', \Diamond \neg a'',\Diamond g, \Diamond \neg g, \Diamond a', \Diamond \neg a', \Diamond c, \Diamond \neg c,  \Diamond h, \Diamond \neg h, \Diamond d, \Diamond \neg d\}  
\end{align*}

Note that the story successfully establishes the unverifiable claim that Solomon is the son of David in the mind of the reader. Given the fact that the reader also entertains the possibility that Solomon might not be David's designated heir (in the face of the multiple verifiable facts around Adonijah's claim), this planted piece of knowledge is the key factor which blocks the introduction of $\Diamond \neg s$ in $T'_4$, and hence prevents the inclusion of the  formula $\Diamond f$ in $T_6$, via the implication $\Diamond \neg s\wedge \Diamond \neg h\to \Diamond f$, which would lead the reader to entertain the possibility that Solomon was part of a foreign coup. 

Indeed, if the episode of Uriah's death was told in a more sanitized  way (e.g.~`Uriah was a valiant soldier who regrettably died in a very bloody battle, and David married his widow out of pity'),  the variables $a$ and $\neg i$ rather than $a$ and $i$ would have been part of $e_4$, and $a$ and $\neg i$ would not have triggered the application of the `exception rule' in the update from $B_3$ to $B_4$, and hence the unverified variable $s$ would have been stored in the form of $\Diamond s$ and $\Diamond\neg s$ in $T'_4$, and hence $\Diamond s$ and $\Diamond\neg s$ would have been added to $T_4$ and persisted to $T_6$, where also $\Diamond \neg h$ would have occurred, leaving the reader free to conclude, via the background formula $\Diamond \neg s\wedge \Diamond \neg h\to \Diamond f$, that $\Diamond f$ is the case, i.e.~that a foreign coup might have occurred. 

\paragraph{Modelling Halpern's analysis using the update protocol for the strategic reader.}

The modelling above captures the conclusions drawn by the audience in the way engineered  by the scribe.  However, to capture  a more critical reader who reasons in the way suggested by Halpern, we apply the update protocol of the strategic listener, as set out in Section \ref{Sec:Dynamically:Formed:Relations}. The set $L'_i$, $L_i$, $B'_i$ and $B_i$ are not affected, nor are $T'_i$ and $T_i$ for $1 \leq i \leq 3$, but from $T'_4$ onwards these sets are different.  

Let us compute the sets $T_i'$. As before, $T'_1 = \{\Diamond g, \Diamond \neg g, \Diamond a', \Diamond \neg a' \}$, since there are no verifiable variables and there are no inconsistencies in $B_1$;
$T'_2 = \{a, \neg e, \Diamond i, \Diamond \neg i, \Diamond a'', \Diamond \neg a'', \Diamond o_1, \Diamond \neg o_1\}$; $T'_3 = \{a,  e, o_2, \Diamond a'', \Diamond \neg a'', \Diamond i, \Diamond \neg i \}$;
the difference is 
$T'_4 = \{a, \neg i, \neg e, o_2, \neg s \}$; notice that we have reversed the polarity of  the `non verifiable' literals $i,  s$. This is in accordance with the idea that  the narrative costs paid by  the writer through the harm to David's reputation are justified by persuading the reader that Salomon is David's biological son. The reader who realizes this, and recognizes this mechanism, should then believe that  the non-verifiable claims made in this episode are most probably false. Then,  $T'_5 = \{a,  e, o_2, \Diamond a'', \Diamond \neg a'', \Diamond i, \Diamond \neg i \}$ and $T'_6 = \{p,j,k,t,m, \Diamond c, \Diamond \neg c, \Diamond h, \Diamond \neg h, \Diamond d, \Diamond \neg d\}$.
We now calculate the sets $T_i$ as follows: 
\begin{align*}
    T_1 &=  \{\Diamond g, \Diamond \neg g, \Diamond a', \Diamond \neg a' \}\\
    T_2 &= \{a, \neg e, \Diamond a'', \Diamond \neg a'', \Diamond i, \Diamond \neg i, \Diamond o_1, \Diamond \neg o_1, \Diamond g, \Diamond \neg g, \Diamond a', \Diamond \neg a'\}\\
    T_3 &= \{a,  o_2, \neg o_1,\Diamond a'', \Diamond \neg a'',  \Diamond i, \Diamond \neg i,   \Diamond g, \Diamond \neg g, \Diamond a', \Diamond \neg a'\}\\
    T_4 &=  \{a,  \neg e, \neg i, o_2, \neg o_1,  \neg s,  \Diamond a'', \Diamond \neg a'',\Diamond g, \Diamond \neg g, \Diamond a', \Diamond \neg a' \}\\
    T_5 &= \{a,  \neg i, o_2, \neg o_1, \neg  s, \Diamond a'', \Diamond \neg a'', \Diamond g, \Diamond \neg g, \Diamond a', \Diamond \neg a' \}\\
    T_6 &= \{a, \neg i, o_2, \neg o_1, \neg s,  p,j,k,t,m, \Diamond a'', \Diamond \neg a'', \Diamond g, \Diamond \neg g, \Diamond a', \Diamond \neg a', \Diamond c, \Diamond \neg c,  \Diamond h, \Diamond \neg h, \Diamond d, \Diamond \neg d, \Diamond f\}  
\end{align*}
Notice now that having $\neg s$ in $T_4$ ensures that it persists into $T_6$  and allows the reader to conclude $\Diamond f$ in $T_6$, leaving open the very damaging possibility, inimical to the writer's aims, that Salomon rose to power through a foreign coup.

\section{Conclusions and further directions}
\label{sec: conclusions}

\paragraph{Main contributions.} In this paper, we have developed a general framework for modelling the extraction of information from strategically biased communication, grounded in the Tiglath-Pileser principle. Building on the codex-based models of persuasion introduced by Rubinstein and Glazer, we shifted the focus from codices to their associated guidance relations, and further generalized these to the broader class of Tiglath-Pileser relations. We provided a set-theoretic and axiomatic analysis of these relations, identifying the structural conditions under which they support successful applications of the Tiglath-Pileser principle in persuasion settings. Within this generalized framework, we reformulated Rubinstein and Glazer's models of  persuasion  and  magical implementation in relational terms, thereby extending their scope beyond one-shot interactions with common knowledge of agents’ goals. Crucially, we introduced a dynamic perspective which accounts for how  Tiglath-Pileser relations  are formed, revised, and strategically manipulated over the course of interaction, giving rise to higher-order reasoning about agents’ strategies and reliability. Finally, we applied this framework to the case studies of the `small businessman and the banker', and Baruch Halpern's historiographical analysis of King David's apology, showing how the proposed machinery can trace the inferential steps from biased narratives to plausible reconstructions of underlying states of affairs.
Several directions for further research naturally emerge from the present framework, which are listed below.

\paragraph{A dynamic account of magical implementation.} In this paper, we developed the dynamic account of persuasion scenarios similar to those treated in \cite{glazer2012model} but we did not develop this approach for magical implementation \cite{glazer2024magical} to the same extent. 
Many real world situations concerning a principal and two or more agents with (partially) conflicting goals, such as judicial trials, proceed in repeated rounds and evolve dynamically. Formalising such situations would require the parties involved to deduce the motivations and strategies of others in real time, which would call for a dynamic version of the magical implementation approach. This is an ongoing line of research. 

\paragraph{Equilibrium analysis.}  Another line of investigation concerns equilibrium analysis. In the present work, Tiglath-Pileser relations are taken as objects that agents form and revise, but their emergence is not yet embedded in a fully strategic model. Integrating the framework into a game-theoretic setting in which agents choose communication strategies while anticipating how their messages will be interpreted via evolving Tiglath-Pileser relations would allow one to study equilibrium notions in which both communication and interpretation co-evolve. This could also be relevant to information extraction and reasoning in environments with competing or adversarial sources.

\paragraph{Probabilistic approaches.} In connection with the game-theoretic approach mentioned above, it would be of interest to develop frameworks in which players' information about the scenario and other players' motivations and strategies can be probabilistic. This would also allow for a quantitative account of how players deal with uncertainty while aggregating information extracted from biased reports, and under which conditions such aggregation converges to accurate beliefs.

\paragraph{Inferring Tiglath-Pileser relations.} In the present paper we proposed mechanisms for inferring the Tiglath-Pileser relations of other agents from repeated observations of their behaviour. In situations where large amounts of data on agents' behaviour are available (e.g.~books or articles of authors, manifestos of political parties, rulings in legal cases), it would be of interest to develop data-analytic methods to extract (approximations of) Tiglath-Pileser relations automatically. This line of inquiry would also be relevant to a research programme seeking to automate the identification of bias or deliberate disinformation and the extraction of reliable information.

\bibliographystyle{siam}
\bibliography{ref}

\begin{thebibliography}{10}

\bibitem{amodei2016concrete}
{\sc D.~Amodei, C.~Olah, J.~Steinhardt, P.~Christiano, J.~Schulman, and D.~Man{\'e}}, {\em Concrete problems in {AI} safety}, arXiv preprint arXiv:1606.06565,  (2016).

\bibitem{atakan2012reputation}
{\sc A.~E. Atakan and M.~Ekmekci}, {\em Reputation in long-run relationships}, The Review of Economic Studies, 79 (2012), pp.~451--480.

\bibitem{bai2022constitutional}
{\sc Y.~Bai, S.~Kadavath, S.~Kundu, A.~Askell, J.~Kernion, A.~Jones, A.~Chen, A.~Goldie, A.~Mirhoseini, C.~McKinnon, C.~Chen, C.~Olsson, C.~Olah, D.~Hernandez, D.~Drain, D.~Ganguli, D.~Li, E.~Tran-Johnson, E.~Perez, J.~Kerr, J.~Mueller, J.~Ladish, J.~Landau, K.~Ndousse, K.~Lukosuite, L.~Lovitt, M.~Sellitto, N.~Elhage, N.~Schiefer, N.~Mercado, N.~DasSarma, R.~Lasenby, R.~Larson, S.~Ringer, S.~Johnston, S.~Kravec, S.~Sheer El~Showk, S.~Fort, T.~Lanham, T.~Telleen-Lawton, T.~Conerly, T.~Henighan, T.~Hume, S.~R. Bowman, Z.~Hatfield-Dodds, B.~Mann, D.~Amodei, N.~Joseph, S.~McCandlish, T.~Brown, and J.~Kaplan}, {\em Constitutional {AI}: Harmlessness from {AI} feedback}, arXiv preprint arXiv:2212.08073,  (2022).

\bibitem{camerer2004cognitive}
{\sc C.~F. Camerer, T.-H. Ho, and J.-K. Chong}, {\em A cognitive hierarchy model of games}, The quarterly journal of economics, 119 (2004), pp.~861--898.

\bibitem{christiano2017deep}
{\sc P.~F. Christiano, J.~Leike, T.~Brown, M.~Martic, S.~Legg, and D.~Amodei}, {\em Deep reinforcement learning from human preferences}, Advances in neural information processing systems, 30 (2017).

\bibitem{conradie2019probabilistic}
{\sc W.~Conradie, S.~Frittella, A.~Palmigiano, A.~Tzimoulis, and N.~Wijnberg}, {\em Probabilistic epistemic updates on algebras}, ACM Transactions on Computational Logic (TOCL), 20 (2019), pp.~1--75.

\bibitem{fagin2004reasoning}
{\sc R.~Fagin, J.~Y. Halpern, Y.~Moses, and M.~Vardi}, {\em Reasoning about knowledge}, 2004.

\bibitem{glazer2012model}
{\sc J.~Glazer and A.~Rubinstein}, {\em A model of persuasion with boundedly rational agents}, Journal of Political Economy, 120 (2012), pp.~1057--1082.

\bibitem{glazer2014complex}
\leavevmode\vrule height 2pt depth -1.6pt width 23pt, {\em Complex questionnaires}, Econometrica, 82 (2014), pp.~1529--1541.

\bibitem{glazer2024magical}
\leavevmode\vrule height 2pt depth -1.6pt width 23pt, {\em Magical implementation},  (2024).

\bibitem{halpern2003david}
{\sc B.~Halpern}, {\em David's secret demons: messiah, murderer, traitor, king}, Wm. B. Eerdmans Publishing, 2003.

\bibitem{KREPS1982253}
{\sc D.~M. Kreps and R.~Wilson}, {\em Reputation and imperfect information}, Journal of Economic Theory, 27 (1982), pp.~253--279.

\bibitem{nagel1995unraveling}
{\sc R.~Nagel}, {\em Unraveling in guessing games: An experimental study}, The American economic review, 85 (1995), pp.~1313--1326.

\bibitem{ross1973economic}
{\sc S.~A. Ross}, {\em The economic theory of agency: The principal's problem}, The American economic review, 63 (1973), pp.~134--139.

\bibitem{stahl1994experimental}
{\sc D.~O. Stahl~II and P.~W. Wilson}, {\em Experimental evidence on players' models of other players}, Journal of economic behavior \& organization, 25 (1994), pp.~309--327.

\bibitem{wardle2017information}
{\sc C.~Wardle and H.~Derakhshan}, {\em {Information disorder: Toward an interdisciplinary framework for research and policymaking}}, vol.~27, Council of Europe Strasbourg, 2017.

\end{thebibliography}
\end{document}